\documentclass[twocolumn,trackchanges]{aastex62}
\usepackage{amsmath,amstext}
\usepackage[T1]{fontenc}
\usepackage{natbib}
\received{date}
\revised{date}
\accepted{date}
\submitjournal{AJ}

\shorttitle{A cold dust-rich debris disk around CP$-$72~2713}
\shortauthors{Mo\'or et al.}

\begin{document}

\title{The big sibling of AU\,Mic: a cold dust-rich debris disk around CP$-$72~2713 in the $\beta$\,Pic moving group}

\correspondingauthor{Attila Mo\'or}
\email{moor@konkoly.hu}

\author{Attila Mo\'or}
\affil{Konkoly Observatory, Research Centre for Astronomy and
Earth Sciences, Konkoly-Thege Mikl\'os \'ut 15-17, H-1121 Budapest, Hungary}
\affil{ELTE E\"otv\"os Lor\'and University, Institute of Physics, P\'azm\'any 
P\'eter s\'et\'any 1/A, 1117 Budapest, Hungary}

\author{Nicole Pawellek} 
\affil{Institute of Astronomy, University of Cambridge, Madingley Road, Cambridge CB3 0HA, UK}
\affil{Max Planck Institute for
Astronomy, K\"onigstuhl 17, D-69117 Heidelberg, Germany}
\affil{Konkoly Observatory, Research Centre for Astronomy and
Earth Sciences, Konkoly-Thege Mikl\'os \'ut 15-17, H-1121 Budapest, Hungary}

\author{P\'eter \'Abrah\'am} 
\affil{Konkoly Observatory, Research Centre for Astronomy and 
Earth Sciences, Konkoly-Thege Mikl\'os \'ut 15-17, H-1121 Budapest, Hungary}
\affil{ELTE E\"otv\"os Lor\'and University, Institute of Physics, P\'azm\'any 
P\'eter s\'et\'any 1/A, 1117 Budapest, Hungary}

\author{\'Agnes K\'osp\'al} \affil{Konkoly Observatory, Research
Centre for Astronomy and Earth Sciences, Konkoly-Thege Mikl\'os \'ut 15-17, H-1121 Budapest, Hungary}
\affil{Max Planck Institute for
Astronomy, K\"onigstuhl 17, D-69117 Heidelberg, Germany}
\affil{ELTE E\"otv\"os Lor\'and University, Institute of Physics, P\'azm\'any 
P\'eter s\'et\'any 1/A, 1117 Budapest, Hungary}

\author{Kriszti\'an Vida} \affil{Konkoly Observatory,
Research Centre for Astronomy and Earth Sciences, Konkoly-Thege Mikl\'os \'ut 15-17, H-1121 Budapest,
Hungary}
\affil{ELTE E\"otv\"os Lor\'and University, Institute of Physics, P\'azm\'any 
P\'eter s\'et\'any 1/A, 1117 Budapest, Hungary}

\author{Andr\'as P\'al} \affil{Konkoly Observatory, Research Centre for Astronomy and
Earth Sciences, Konkoly-Thege
Mikl\'os \'ut 15-17, H-1121 Budapest, Hungary}
\affil{ELTE E\"otv\"os Lor\'and University, Institute of Physics, P\'azm\'any 
P\'eter s\'et\'any 1/A, 1117 Budapest, Hungary}

\author{Anne Dutrey} \affil{Laboratoire d'Astrophysique de Bordeaux, 
Univ. Bordeaux, CNRS, B18N, All\'ee Geoffroy Saint-Hilaire, F-33615 Pessac, France}

\author{Emmanuel Di Folco} \affil{Laboratoire d'Astrophysique 
de Bordeaux, Univ. Bordeaux, CNRS, B18N, All\'ee Geoffroy Saint-Hilaire, F-33615 Pessac, France}

\author{A. Meredith Hughes} \affil{Department of Astronomy, Van
Vleck Observatory, Wesleyan University, 96 Foss Hill Drive,
Middletown, CT 06459, USA}

\author{Quentin Kral} \affil{LESIA, Observatoire de Paris, Universit\'e PSL, CNRS, Sorbonne Universit\'e, Univ. 
Paris Diderot, Sorbonne Paris Cit\'e, 5 place Jules Janssen, F-92195 Meudon, France}

\author{Ilaria Pascucci} \affil{Lunar and Planetary Laboratory, The University of 
Arizona, Tucson, AZ 85721, USA}

\begin{abstract} 
Analyzing {\sl Spitzer} and {\sl Herschel} archival measurements we identified a 
debris disk around the young K7/M0 star CP$-$72~2713. The system belongs to the 24\,Myr old $\beta$~Pic 
moving group. Our new 1.33\,mm continuum observation, obtained with the ALMA 7-m array, 
revealed an extended dust disk with a peak radius of 140\,au, probably tracing the location 
of the planetesimal belt in the system. The disk is outstandingly large compared to known 
spatially resolved debris disks and also to protoplanetary disks around stars of comparable
masses. 
{The dynamical excitation of the belt at this radius is found to 
be reconcilable with planetary stirring, while self-stirring by large planetesimals 
embedded in the belt can work only if these bodies 
form very rapidly, e.g. via pebble concentration.}
By analyzing the 
spectral energy distribution we derived a characteristic dust temperature of 43\,K and a 
fractional luminosity of 1.1$\times$10$^{-3}$. 
 The latter value is prominently high, we know of only four 
other similarly dust-rich Kuiper-belt analogs within 40\,pc of the Sun.

\end{abstract} 

  
\section{Introduction} \label{sec:intro}
Following the dispersal of their gas-rich primordial disks, typically by $\sim$10\,Myr 
\citep[e.g.][]{ercolano2017}, young stars are thought to be 
surrounded by planetesimals not incorporated into planets. 
Depending on the local level 
of dynamical excitation, in
certain regions of the circumstellar environment low-velocity collisions between
planetesimals can result in mergers, while in more stirred regions
fragmentation can occur \citep{wyatt2008}. Second generation debris dust grains, 
produced in the emerging collisional cascade \citep{hughes2018}, then make the sufficiently stirred
parts of the disk observable through their thermal emission at infrared (IR) and millimeter 
wavelengths, and in optical/near-IR scattered light. Studying the spatial 
distribution of debris material can thus reveal regions where planetesimals could form but 
where the buildup of planets was halted by some environmental 
conditions. 

Knowledge about the location of dust producing planetesimal belts 
in young systems 
can also allow us to constrain
our models on the possible dynamical excitation mechanisms \citep{mustill2009} as well as 
pinpointing how these 
belts form \citep{matra2018}. 
Further, the last few years have seen the discovery of an increasing number of young 
debris disks with a detectable amount of gas material \citep[e.g.][]{moor2011,lieman-sifry2016}. 
In most cases this gas is likely second generation \citep{kral2019},
originated from erosional processes. It implies the presence of
volatile-rich bodies \citep{kral2017}, thus allowing us to probe the
ice abundances of (exo-)planetesimals \citep{kral2016,matra2017,matra2019}.

A significant fraction of the known young stars (10--50\,Myr) 
in our neighbourhood belong to gravitationally unbound, loose associations of 
stars called moving groups \citep{torres2008}. Previous IR observations
obtained with spaceborne telescopes led to the discovery of 
many debris disks, 
including some of the most 
iconic ones \citep[e.g. $\beta$\,Pic, AU\,Mic,][]{aumann1985,mathioudakis1991}, 
around the members of these kinematic 
assemblages \citep[e.g.][]{rebull2008,zuckerman2011,donaldson2012,rm2014,moor2016}. 
These studies typically revealed higher excess detection rates in moving groups 
than in samples of older field stars which contain on average less dust 
as a result of longer collisional evolution of their planetesimal belts. 
This is particularly noticeable for debris disks around stars with spectral types later than 
K5, where the majority of the small number of known systems are harbored by 
moving group members 
\citep[e.g.,][]{kalas2004,low2005,olofsson2018,flaherty2019,sissa2018,zuckerman2019}.
Though the detection rate for debris disks with late type hosts 
is higher in young groups than among field stars, it is probable that 
the IR excess of several late-type members has remained undetected 
because their disks are faint due to the low luminosity of the host stars
and/or their smaller disk masses on average.

By carefully inspecting the archives of far-infrared observations obtained 
by the {\sl Spitzer Space Telescope} and the {\sl Herschel Space Observatory}, 
we identified a cold, dust-rich debris disk around a nearby star CP$-$72~2713 
\citep[36.62$\pm$0.03\,pc,][]{brown2018,lindegren2018,cb2018}\footnote{In the proof 
phase we were informed that the discovery of this debris disk 
has already been announced in a talk \citep{plavchan2009} and studied in 
a recently accepted paper \citep{tanner2020} as well as in an ongoing research project 
(P. Plavchan, in prep.). Thus our present work may be considered as an independent 
rediscovery of the disk.}.
The star is late-type, \citet{torres2006} derived a spectral type of K7e~V from a high resolution 
spectrum, while \citet{pecaut2013} and \citet{gaidos2014} obtained spectral types of K7e~IV and M0, respectively, 
based on low resolution spectra.  
As a member of  
the $\sim$24\,Myr old $\beta$\,Pic 
moving group \citep[BPMG,][]{torres2006,bell2015,lee2018,gagne2018}, 
this disk provides a rare opportunity to study the early evolution 
of planetesimal belts surrounding late type stars.

In this paper we present a detailed analysis of this system by using
multiwavelength observations.
Based on spatially resolved far-IR {\sl Herschel}/PACS and millimeter ALMA images 
of the dust emission, we model the structure of the disk and constrain the location 
of the dust producing planetesimal belt. 
Additionally, we carried out ALMA line observations 
to look for CO molecules in the disk.

\section{Observations and Data Reduction} \label{sec:obs}

\subsection{Spitzer/MIPS}
CP$-$72~2713 was observed with the Multiband Imaging Photometer for
Spitzer \citep[MIPS,][]{rieke2004} on-board the Spitzer Space
Telescope \citep{werner2004} at 24{\micron} and 70{\micron} on 2008
November 23 (PI: P. Plavchan). Photometry at 24{\micron} was taken from
the Spitzer Enhanced Imaging Products (SEIP) catalog. 
The photometric uncertainty was computed as the quadratic sum of 
the instrumental noise listed by the catalog and the calibration 
error of 4\% (MIPS Instrument 
Handbook\footnote{\url{https://irsa.ipac.caltech.edu/data/SPITZER/docs/mips/mipsinstrumenthandbook/}}).
The SEIP
catalog does not include data from the 70{\micron} channel. Therefore,
we downloaded the pipeline reduced Basic Calibrated Data images
(version S18.13.0) from the Spitzer Heritage Archive and used MOPEX
\citep[MOsaicking and Point source Extraction,][]{makovoz2005} to
co-add them and to correct for array distortions. 
For further 
improvement prior to the co-addition we applied column mean subtraction and time 
filtering for the images following \citet{gordon2007}.
The final image shows a point source close to the Gaia\,DR2 position of CP$-$72~2713 
(corrected for proper motion between the epochs of observations). The separation 
is only 0\farcs9 ($\Delta RA = -0\farcs86\pm0\farcs60$, $\Delta DEC = +0\farcs16\pm0\farcs60$), 
that is within the 1$\sigma$ uncertainty (1\farcs7) of the pointing 
reconstruction at 70{\micron} (MIPS Instrument Handbook). It is
somewhat smaller than the typical angular offsets measured by \citet[][see their fig.~5]{carpenter2008} 
 and \citet[][fig.~3]{moor2011b} in MIPS 70{\micron} observations 
of debris disks with similar or higher signal-to-noise ratios (SNR$>$25). 
Therefore, it can be concluded that the observed emission matches well 
the position of our target.

We performed aperture photometry for this source with a radius of
16{\arcsec}, and sky annulus between 39{\arcsec} and 65{\arcsec}. We applied the
proper aperture correction factor and adopted a calibration error of
7\% (MIPS Instrument Handbook) that was added quadratically to the
measured uncertainty. The MIPS flux densities are given in Table~\ref{phottable}.

\subsection{Herschel/PACS Observations} \label{sec:pacsobs}
We processed far-infrared imaging observations of CP$-$72 2713 carried
out with the Photodetector Array Camera and Spectrometer 
\citep[PACS,][]{poglitsch2010} instrument of the {\sl Herschel Space Observatory} on
2013 April 23 (PI: A. Tanner). These measurements were performed in
mini scan-map mode (PACS Observer's Manual v2.5.13) at 100 and
160{\micron}. The data reduction and calibration were done in the
Herschel Interactive Processing Environment (HIPE, Ott 2010) version
14.2 using the standard pipeline script optimized for mini scan-map
observations and applying the PACS calibration tree No. 78. To
mitigate the low-frequency (1/f) noise present in the data we applied
highpass filtering with filter width parameters of 20 and 35 at 100
and 160{\micron}, respectively. Since this process introduces flux
loss, the immediate vicinity of the target was excluded from the
filtering using a 25$''$ radius circular mask positioned at the
source's location. The final images were generated with pixel sizes of
1{\arcsec} at 100{\micron} and 2{\arcsec} at 160{\micron}.

\begin{figure*}
\begin{center}
\includegraphics[angle=0,scale=.40,bb=0 22 372 452]{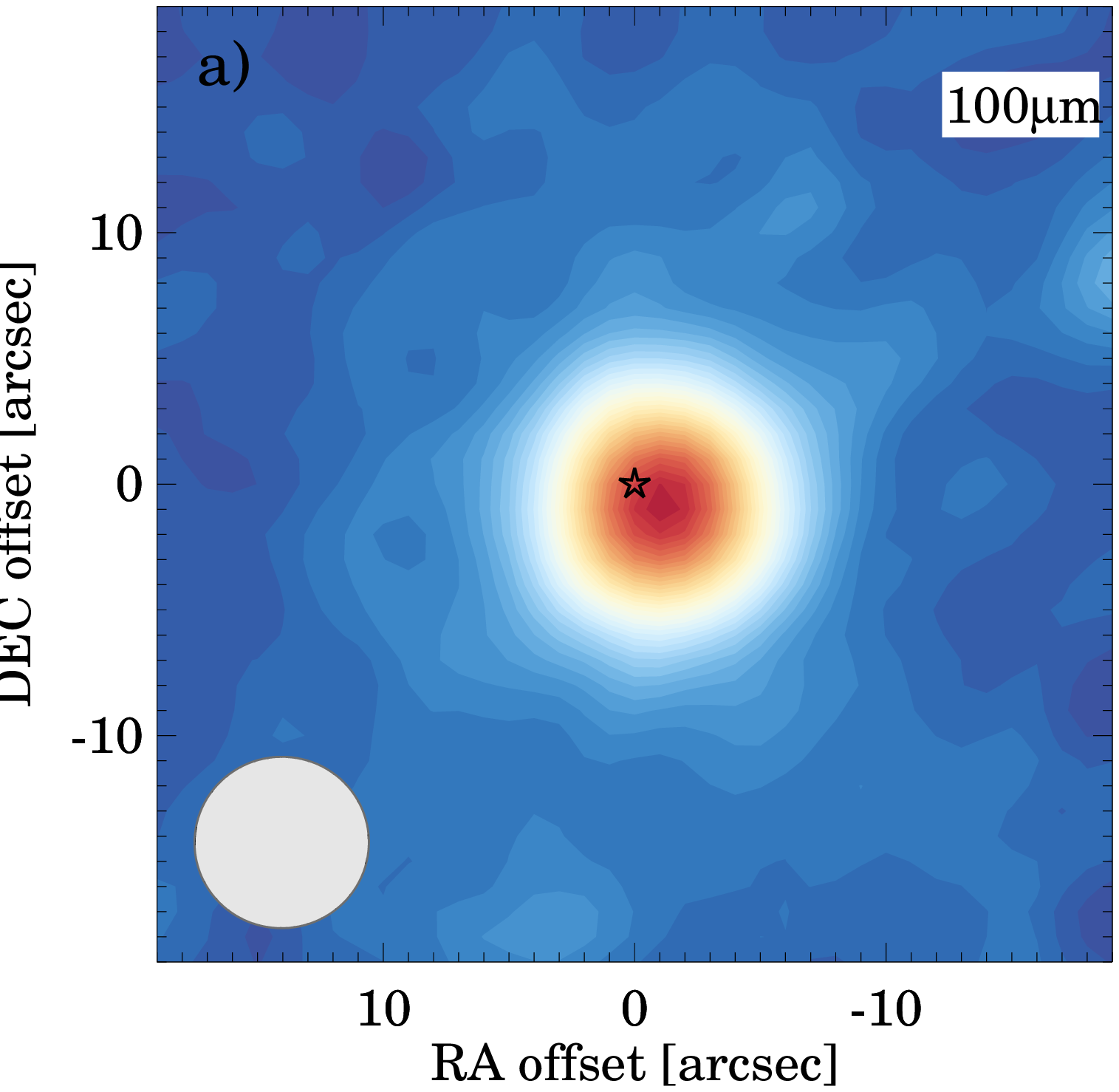}
\includegraphics[angle=0,scale=.40,bb=0 22 372 452]{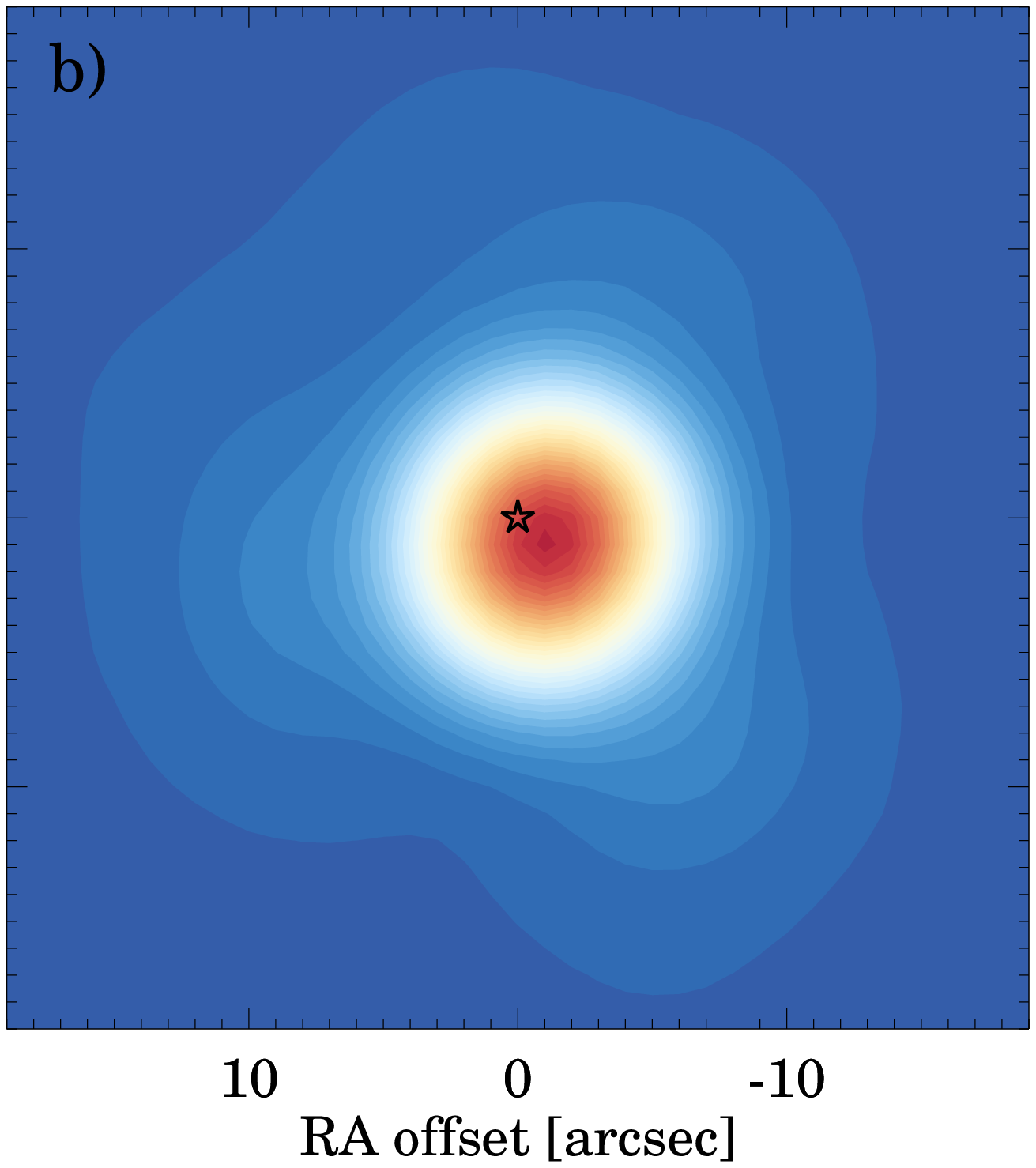}
\includegraphics[angle=0,scale=.40,bb=0 22 511 452]{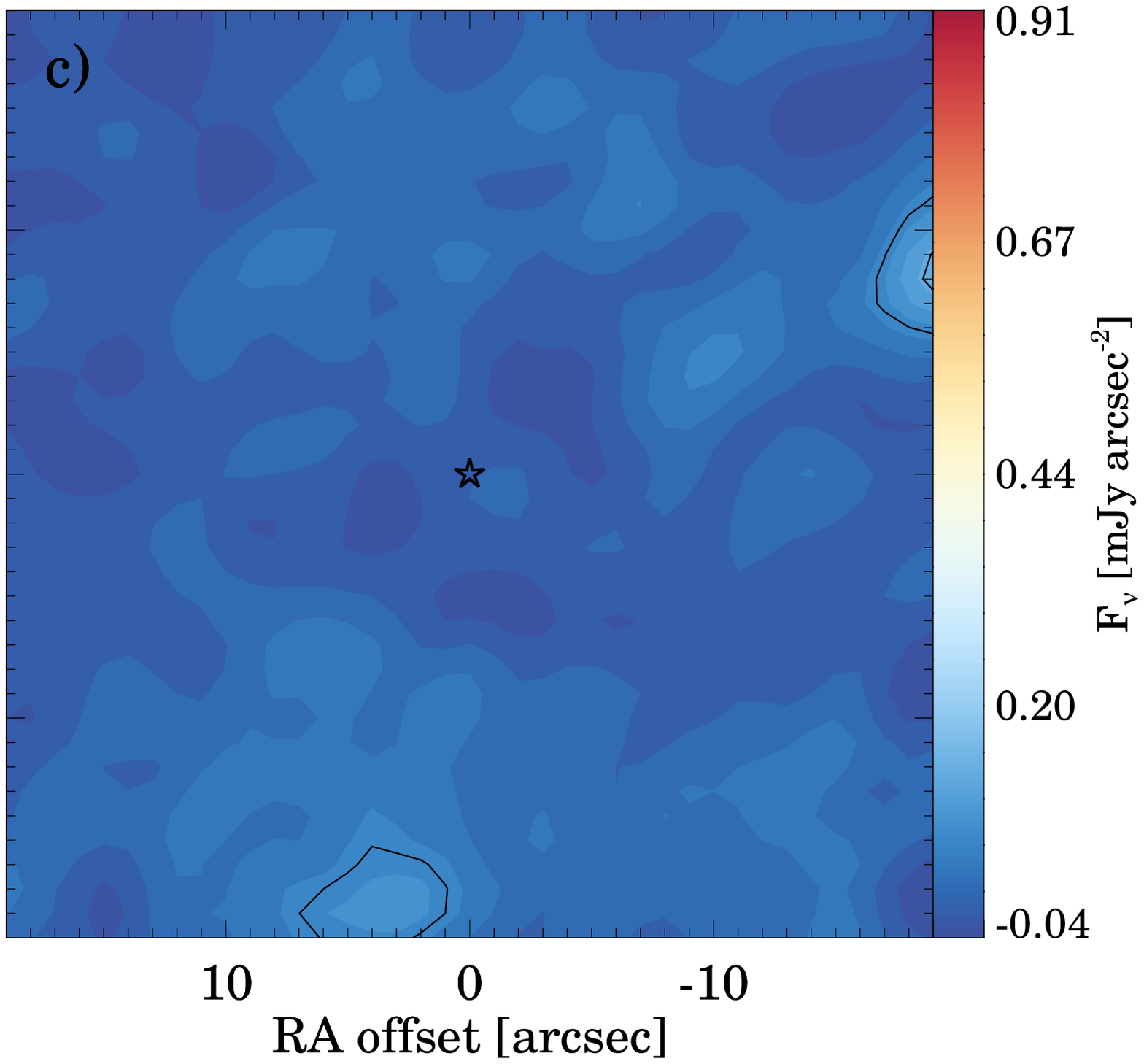}
\includegraphics[angle=0,scale=.40,bb=0 22 372 452]{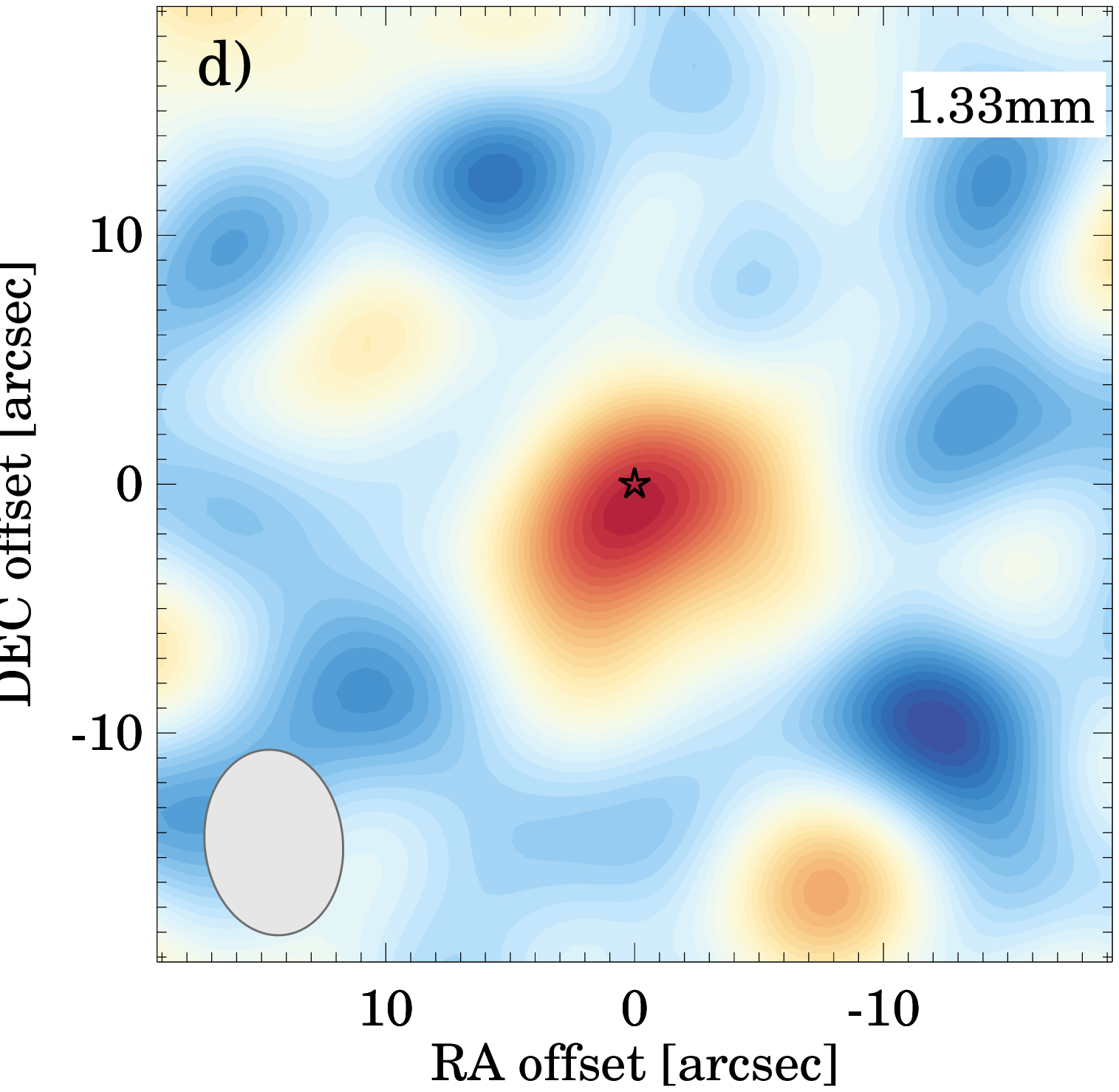}
\includegraphics[angle=0,scale=.40,bb=0 22 372 452]{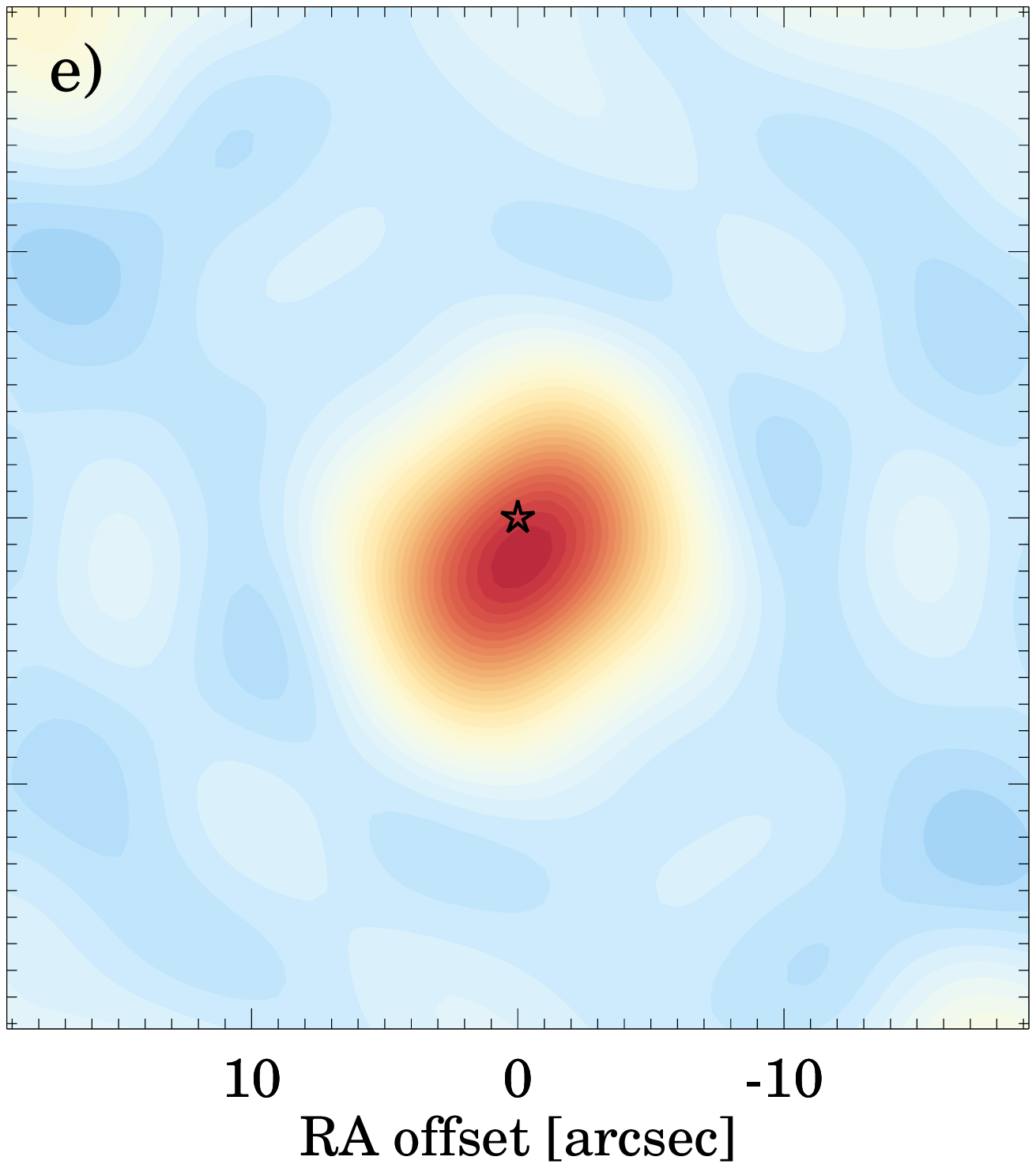}
\includegraphics[angle=0,scale=.40,bb=0 22 511 452]{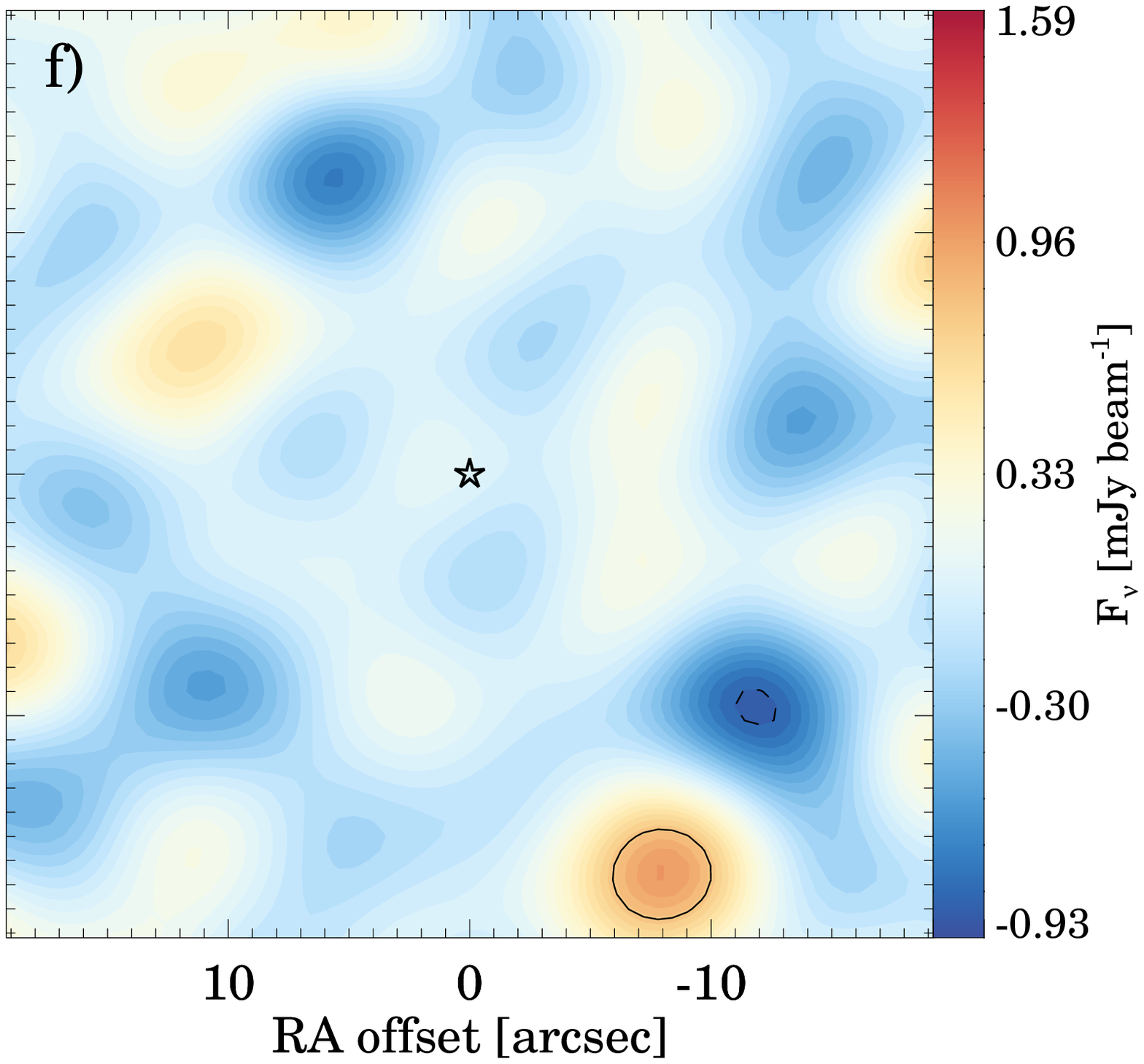}
\end{center}
\caption{Upper panels: (a) PACS 100{\micron} image of CP$-$72~2713, (b) simulated image of the 
best fit dust ring model convolved with our PACS PSF (see Sect~\ref{sec:pacsobs}), 
(c) the image of residuals.  The stellar position is marked 
with a black star symbol. The gray filled ellipse in the lower-left corner represents the size and shape 
of the PACS PSF (the FWHMs and position angle of the ellipse were obtained by fitting a 2-dimensional 
Gaussian to the 100{\micron} image of $\alpha$~Boo, see Sect.~\ref{sec:obs}). 
In the residual image, contours of 3$\times$ the rms noise are also shown. 
Lower panels: (d) 1.33\,mm ALMA continuum image, (e) image of the best-fit 
model, and (f) the residuals image. We note that our modelling was performed in the uv space 
(Sect.~\ref{sec:obs}). 
The (e) and (f) panels display CLEAN images of the best-fit model and residuals after
subtracting the model visibilities from the ALMA observations using Briggs weighting with 
a robustness parameter of 0.5 in the imaging. In (f) panel the dashed and solid contours 
correspond to the --3$\sigma$ and +3$\sigma$ levels.
\label{fig:images}}
\end{figure*}

CP$-$72 2713 was clearly detected in both PACS bands. The 100{\micron}
image is displayed in Fig.~\ref{fig:images}a. 
The positional offsets between the
source's centroid and its proper motion corrected Gaia DR2 position is
1$\farcs$4 ($\Delta RA = -1\farcs06\pm0\farcs07$, $\Delta DEC = -0\farcs84\pm0\farcs07$) and 
1$\farcs$6 ($\Delta RA = -1\farcs35\pm0\farcs20$, $\Delta DEC = -0\farcs76\pm0\farcs20$) 
at 100{\micron} and 160{\micron}, respectively.
\citet{sportal2014} derived a typical absolute pointing accuracy of 0\farcs9 for the
phase of the {\sl Herschel} mission when our target's observations were performed.
Utilizing large number of PACS mini scan-maps of standard stars, \citet{marton2017}
found a median positional difference of 1$\farcs$5 for the 
blue PACS detector (70{\micron} and 100{\micron} filters) and 1$\farcs$7 for the red PACS 
detector (160{\micron}).
Our measured offsets correspond well to the latter values implying their instrumental origin.

We measured the
source's flux by placing a 15{\arcsec} radius aperture onto the centroid
position, and a sky annulus between 60{\arcsec} and 70{\arcsec}. Aperture
corrections were taken from the appropriate calibration files. To
estimate the uncertainty of the measured flux densities we placed
sixteen apertures with radii identical to the source aperture evenly
along the sky annulus and carried out photometry without background
subtraction in each of them. Then the uncertainty was derived as the
standard deviation of these background flux values. The final
photometric uncertainty was calculated as a quadratic sum of this and
the absolute calibrational uncertainty of the PACS detector 
\citep[7\%,][]{balog2014}. The resulting flux densities are in Table~\ref{phottable}.

To evaluate whether the detected source is spatially extended we
fitted a 2-dimensional Gaussian model to the images using the
\texttt{mpfit2dpeak} IDL procedure \citep{markwardt2009}. At 100{\micron} the best
fit Gaussian has a Full Width at Half Maximum (FWHM) of
8$\farcs$5$\pm$0$\farcs$3$\times$8$\farcs$1$\pm$0$\farcs$3, while the
FWHM sizes of the point spread function (PSF) in maps performed with
scan speed of 20{\arcsec}/s (as in our case) are
6$\farcs$89$\times$6$\farcs$69 (PACS Observer's Manual). This implies
that CP$-$72~2713 is marginally resolved at 100{\micron}. The fitted
Gaussian at 160{\micron} is consistent with the PSF.

Assuming that the observed emission arises from a circumstellar dust
ring, we fitted a grid of simple disk models to the 100{\micron} image.
The models have three free parameters: the radius of the ring ($R$),
the position angle ($PA$, measured east of north), and the inclination
($i$, angle of the disk plane with respect to the sky plane). 
Considering that the observation 
does not resolve the radial structure we assumed a Gaussian 
radial surface brightness profile that peaks at $R$ and has an FWHM fixed to 0.2$R$.
Since at this wavelength the stellar
photosphere is predicted to be $\sim$100$\times$ fainter than the
measured flux density (Sect.~\ref{sec:results}) we neglected its contribution in our
modelling. Disk images constructed in this way were then convolved
with a PSF model that we constructed from mini-scan map observations
of $\alpha$ Boo (OBS ID 1342247702/1342247703), processed in the same
way as that of our target. The wings of the PACS beam form a
characteristic tri-lobe pattern whose orientation depends on the
actual roll angle of the observation. To allow direct comparison, we
rotated the image of $\alpha$ Boo to match the telescope's roll angle
at the observation of CP$-$72~2713. In order to select the best-fitting
model and estimate the uncertainties of the fitted parameters, we used
Bayesian inference following \citet{moor2015}. The best-fitting
model parameters are: $R=2\farcs7^{+0\farcs7}_{-0\farcs6}$ (or
99$^{+22}_{-26}$\,au), $i = 29{\degr}^{+26{\degr}}_{-21{\degr}}$, 
$PA = 153{\degr}^{+60{\degr}}_{-64{\degr}}$. This model and the
residuals are displayed in Figure~\ref{fig:images}b-c. As Fig.~\ref{fig:images}c shows, toward the 
source no residuals higher than 3$\sigma$ were observed. The $>$3$\sigma$ residuals at the
edge of the displayed image are at an angular separation $>$15{\arcsec} and may be related to 
background sources.

\subsection{Observations with the ALMA 7-m Array} \label{sec:almaobs}
We observed CP$-$72~2713 with the ALMA 7-m array in the framework of a 
Cycle 4 project (2016.2.00200.S, PI: Á. Kóspál) that focused on the gas content of 
young dust-rich debris disks. Our Band~6 observations 
include two 1.875\,GHz wide continuum spectral windows each with 128 channels 
centered at 217.0 and 233.5\,GHz and two additional 
windows that were designed to cover the J = 2--1 rotational lines of $^{12}$CO, $^{13}$CO, 
and C$^{18}$O. The latter two lines were measured in a single window using a channel width 
of 428.28\,kHz and thereby providing a bandwidth of 1.875\,GHz with 4096 channels. In the 
case of $^{12}$CO the channel width was set to 122.07\,kHz, this band had a total width of 
468.8\,MHz. Two observations were performed, one day apart from each other, on 2017 July 25 
and 26. In both cases nine antennas were involved providing projected baseline lengths ranging 
between 8.9 and 44.7 m (6.7--33.6\,k$\lambda$). The obtained data were calibrated and flagged 
with the ALMA reduction tool  {\sl Common Astronomy Software Applications} 
\citep[CASA v4.7.2;][]{mcmullin2007} utilising the pipeline script delivered with the data.

\paragraph{Continuum emission}
We used the \texttt{tclean} task in CASA \citep[v5.6.2,][]{mcmullin2007} to construct the 
continuum image by applying Briggs weighting with a robustness parameter of 0.5. Since no 
line was detected (see below) we combined non-flagged data from all spectral windows achieving 
an rms noise level of 0.28mJy~beam$^{-1}$. The beam size is 7\farcs4$\times$5\farcs5 with a 
position angle of 6\fdg3. The obtained 1.33\,mm continuum image (Figure~\ref{fig:images}d) 
shows a moderately resolved source around CP$-$72~2713 with a peak signal-to-noise ratio (SNR) of 5.7. 
To determine the 
source's flux density and to constrain the spatial distribution of emitting dust grains we 
used the CASA-based \texttt{uvmultifit} tool \citep{marti-vidal2014} that allows to fit models 
to the measured visibilities. Prior to this running, the data weights were recomputed using the 
\texttt{statwt} task. Similarly to the case of PACS image analysis, we adopted a Gaussian ring 
model with a fixed FWHM width of 0.2$R$. The obtained best-fit solution has $R = 140\pm14$\,au 
(3\farcs8$\pm$0\farcs4), $i = 46{\degr}\pm12{\degr}$, $PA = 130{\degr}\pm16{\degr}$ and yields 
a flux density of 3.8$\pm$0.6\,mJy (the quoted uncertainty includes a 10\% calibration 
uncertainty). We used the \texttt{tclean} task to compile images of the best-fit model and 
the residuals (Figure~\ref{fig:images}e-f). No significant residual emission appears in the 
disk region. The fitted center of the model ring shows an offset of 1\farcs3 from the stellar 
position ($\Delta RA = -0\farcs3\pm0\farcs42$, $\Delta DEC = -1\farcs22\pm0\farcs45$). 
The positional accuracy of an ALMA observation with normal calibration can be estimated as the ratio of 
the achieved resolution and SNR, i.e. $\sim$0\farcs9 in our case, thus the observed offset 
is not significant. The best-fitting disk parameters
derived from the analysis of PACS and ALMA data are summarized in Table~\ref{tab:disk}.

\begin{deluxetable}{ccccc}[h!]                                                      
\tabletypesize{\scriptsize}                                                     
\tablecaption{Measured continuum fluxes and predicted photospheric fluxes for CP$-$72~2713 \label{phottable}}
\tablewidth{0pt}                                                                
\tablecolumns{5}                                                                
\tablehead{ \colhead{$\lambda$} & \colhead{Meas. flux$^{a}$} &                  
\colhead{Instrument} &  \colhead{Photosph. flux} &  \colhead{Reference} \\              
\colhead{({\micron})} & \colhead{(mJy)} &                                       
\colhead{} &  \colhead{(mJy)} &  \colhead{}                                     
}
%\decimalcolnumbers                                                                               
\startdata                                                                      
        9.00 &         128.1$\pm$12.3 &                IRC & 105.2 & \citet{ishihara2010} \\
       11.56 &           60.0$\pm$2.8 &               WISE & 65.0 &    \citet{cutri2013} \\
       22.09 &           18.9$\pm$1.5 &               WISE & 18.3 &    \citet{cutri2013} \\
       23.68 &           16.1$\pm$0.6 &               MIPS & 16.0 &                 SEIP \\
       71.42 &           71.1$\pm$5.6 &               MIPS & 1.8 &            This work \\
  100$^b$ &          110.6$\pm$9.7 &               PACS & 0.9 &            This work \\
      160 &         103.9$\pm$16.9 &               PACS & 0.35 &            This work \\
 1330$^b$ &          3.80$\pm$0.59 &               ALMA & 0.005 &            This work \\
 8820     &         0.096$\pm$0.023&               ATCA & 10.6$\times10^{-5}$  &   Norfolk et al. in prep. \\          
\enddata                                                                        
\tablenotetext{a}{The quoted flux densities are not color corrected.}           
\tablenotetext{b}{The emission is marginally resolved at these wavelengths.}    
\end{deluxetable}

\paragraph{CO line data}
We used CASA to compile the spectral line cubes. 
After the continuum emission was subtracted from the calibrated visibilities using the 
\texttt{uvcontsub} task, the imaging was done with the \texttt{tclean} task. 
In the latter step we used Briggs weighting with a robustness parameter of 0.5 and 
the channel width was set to 0.4\,km~s$^{-1}$ for the $^{12}$CO line and to 
 1.4\,km~s$^{-1}$ for data of the rarer $^{13}$CO and C$^{18}$O isotopologues. 
Using the radial velocity data from Table~\ref{tab:stellar} 
we derived a radial velocity of $+$0.4\,km~s$^{-1}$ for CP$-$72~2713 in the LSR frame. We found 
no significant emission at this velocity (or anywhere else) in 
the CO data cubes. 

In order to estimate an upper limit for the $^{12}$CO (2--1) integrated line flux   
we assumed that the possible CO gas is co-located with the dust material in the disk.
We defined a disk mask considering those regions in the millimeter continuum image that are 
adjacent with the maximum brightness and having SNR$>$2. Using the obtained mask we constructed
the spectrum and calculated its rms. To consider the correlation between the neighboring channels due to 
the applied weighting function, this rms was multiplied by $\sqrt{2.667}$ \citep[see][and references therein]{matra2019}.
Since based on our results 
the bulk of the dust is located at radii $\gtrsim$100\,au and the disk inclination is 
$\lesssim$82{\degr} (3$\sigma$ confidence interval), by adopting a stellar mass of 
0.71\,M$_\odot$ (see Sect.~\ref{sec:stellarprops})
we expect a maximum line width of 5~km~s$^{-1}$. Using the corrected rms value 
and this linewidth 
we derived a 3$\sigma$ upper limit of 0.216\,Jy~km~s$^{-1}$ for the
$^{12}$CO (2--1) integrated line flux.

\begin{deluxetable}{lccc}[h!]                                                                                                  
\tablecaption{Best-fitting dust disk parameters derived from the analysis of PACS and ALMA 
data \label{tab:disk}}
\tablewidth{0pt}                                                                
\tablecolumns{4}                                                                
\tablehead{\colhead{Data set} & \colhead{PA (\degr)} & \colhead{i (\degr)} & \colhead{R ({\rm au})}}
%\decimalcolnumbers                                                                               
\startdata                                                                      
PACS 100{\micron}   &  153$^{+60}_{-64}$  & 29$^{+26}_{-21}$ & 99$^{+22}_{-26}$ \\
ALMA 1.33\,mm       &  130$\pm$16         & 46$\pm$12        & 140$\pm$14 \\
\enddata                                                                           
\end{deluxetable}

\section{Results and analysis} \label{sec:results}

\subsection{Basic stellar properties of CP$-$72~2713}\label{sec:stellarprops}
To estimate the basic stellar parameters of CP$-$72~2713 and provide
photospheric flux predictions at infrared and millimeter wavelengths, 
we modeled the stellar photosphere by fitting an ATLAS9 atmosphere
model \citep{castelli} to the photometric data of the star.
CP$-$72~2713 is an active star that exhibits variability due to spots (see Fig.~\ref{fig:tess}). 
Using photometric data from the ASAS survey, \citet{kiraga2012} analyzed 
498 and 191 measurements in the  $V$ and $I$ band, respectively. They
derived average magnitudes and inferred variability amplitudes of 0.16 and 0.017\,mag 
in the $V$ and $I$ bands. 
We derived 
an amplitude of $\sim$0.04 magnitude in the TESS band (Appendix~\ref{appendix}). 
To reduce uncertainties stemming from the spot-induced variability in 
the optical regime, we limited our analysis to data that are based on
longer term monitoring observations taken from \citet{kiraga2012}.
 This was further supplemented 
 by near-IR ($J$, $H$, $K_{\rm s}$) photometry from the Two Micron All-Sky Survey \citep[2MASS,][]{skrutskie2006}, 
 and by $W1$, $W2$ band data from the 
 AllWISE \citep{cutri2013} catalog. This longer wavelength photometry is typically 
 less susceptible to variations caused by spots \citep[e.g.,][]{goulding2012}.    
Based on the recent 3D reddening map of the local interstellar medium, that is available through the \texttt{Stilism} 
tool\footnote{\url{https://stilism.obspm.fr/}} \citep{capitanio2017,lallement2018}, 
the reddening towards our target is negligible. Thus we assumed $E(B-V)=0$ 
in our fitting. 
By adopting solar metallicity and a $\log{g}$ of 4.5\,dex, our $\chi^2$ minimization
yielded an effective temperature of 3900$\pm$70\,K for CP$-$72~2713.
Using the inferred photosphere model and the Gaia DR2 distance of the object we derived 
a luminosity of 0.18$\pm$0.01\,L$_\odot$ and a stellar radius of 
0.94$\pm$0.04\,R$_\odot$. To estimate the stellar mass and age, the effective 
temperature and the obtained luminosity values were compared to the theoretical predictions 
of the MESA Isochrones and Stellar Tracks \citep[MIST,][]{choi2016,dotter2016}. 
In the course of isochrone fitting we followed the method described by 
\citet{pascucci2016}, that yielded 0.71$^{+0.03}_{-0.05}$\,M$_\odot$ and
14$^{+5}_{-3}$\,Myr  for the best-fit mass and age, respectively. 
Considering the uncertainties, the estimated age is satisfactorily consistent with that 
of the BPMG \citep[24$\pm$3\,Myr,][]{bell2015}. 
We note that the derived stellar mass and radius result in a 
$\log{g}$ of 4.34\,dex, adequately confirming our a priori assumption.
The stellar parameters are summarized in Table~\ref{tab:stellar}.

\subsection{Stellar activity} \label{sec:stellaractivity}

Similarly to other young late-type stars, CP$-$72~2713 shows strong stellar 
activity that manifests in various indicators. It has an X-ray 
counterpart in the ROSAT catalog implying a fractional X-ray luminosity 
of $\log{L_{\rm X}/L_{\rm bol}} = -2.92$ \citep{kiraga2012}.
This is close to the saturation value of -3, indicating strong coronal activity.
The star was also detected by the GALEX satellite \citep{martin2005} both in 
the near- ($NUV$) and far-UV ($FUV$) bands. By combining these photometric data 
with 2MASS measurements and placing the object in the 
$NUV-J$ vs. $J-K$ color-color diagram, compiled by \citet{findeisen2011}, 
we found that its position matches well the locus of other BPMG members.
This implies significant excess at UV wavelengths, which is likely 
due to prominent chromospheric activity.

CP$-$72~2713 has already been observed by The Transiting Exoplanet Survey Satellite 
\citep[{\sl TESS},][]{ricker2015} 
providing a very precise, $\sim$52\,days long photometric 
data set with a cadence of 2\,minutes. 
Reduction and analysis of these {\sl TESS} data are described in detail in the Appendix~\ref{appendix}.
 As further signatures of activity, CP$-$72~2713 exhibits 
flares and rotational modulation due to stellar spots in its 
{\sl TESS} light curve (Fig.~\ref{fig:tess}). Consistently with previous ASAS based  
results of 4.456\,days \citep{kiraga2012} and 4.48\,days \citep{messina2010}
we obtained a rotation period of P = 4.437\,d from a Fourier analysis of our light curve. 
In addition, we identified 51 flares or 
group of flares. We found that the star spent about 936\,minutes in a flaring 
state, which is 1.24\% of the time covered by {\sl TESS}.

\begin{deluxetable}{ccc}[h!]                                                      
\tabletypesize{\scriptsize}                                                     
\tablecaption{Stellar parameters \label{tab:stellar}}
\tablewidth{0pt}                                                                
\tablecolumns{3}                                                                
\tablehead{ \colhead{Parameter} & \colhead{Data}& \colhead{References}}                                                              
\startdata
Distance (pc)       &  36.62$\pm$0.03\,pc & \citet{cb2018} \\
Spectral type       &  K7Ve  & \citet{torres2006}  \\
                    &  K7IVe & \citet{pecaut2013} \\
		    &  M0    & \citet{gaidos2014}  \\ 
$T_{\rm eff}$ (K)   &  3900$\pm$70\,K & This work \\
$L_*$  ($L_\odot$)  &  0.18$\pm$0.01 & This work \\
$R_*$  ($R_\odot$)  &  0.94$\pm$0.04 & This work  \\ 
$M_*$  ($M_\odot$)  &  0.71$^{+0.03}_{-0.05}$ & This work  \\
$L_{\rm X}/L_{\rm bol}$ & -2.92 & \citet{kiraga2012}\\
$v\sin{i_*}$ (\,kms$^{-1}$) &  7.1$\pm$0.9$^a$  & \citet{torres2006} \\
                            &                   & \citet{weise2010} \\
$P_{\rm rot}$ (days) & 4.437 & This work \\
$v_{\rm rad}$ (kms$^{-1}$) & 8.0$\pm$0.4$^a$ & \citet{torres2006} \\ 
                                 &                       & \citet{lindegren2018} \\  
\enddata 
\tablenotetext{a}{Weighted average of the values reported by the cited papers.}                                                                  
\end{deluxetable}

\subsection{Disk properties} \label{sec:diskprops}
\begin{figure}[h!] 
\includegraphics[scale=.48,angle=0]{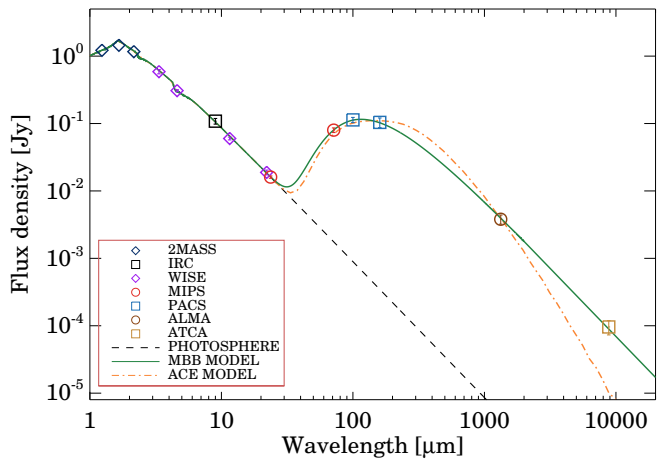}
\caption{Color-corrected SED of CP$-$72 2713. The green curve shows the best 
fitting modified blackbody model. As a comparison, the orange curve shows an Analysis of 
Collisional Evolution (ACE) model for the disk with a total mass of 50\,M$_\oplus$ 
was also overplotted (see Sect.~\ref{sec:ace}). This shows the model SED after 24\,Myr of evolution 
assuming $\dot{M}_{\rm *,wind}= 300 \dot{M}_{\rm \odot,wind}$. 
\label{fig:sedplot}
}
\end{figure}

To construct the spectral energy distribution (SED) of our target at $\lambda>5${\micron}, the already reported {\sl Spitzer}, 
{\sl Herschel} and ALMA data were supplemented by additional spaceborne mid-infrared photometry obtained by the 
{\sl AKARI} and {\sl WISE} 
satellites. The SED was further supplemented by a recent observation obtained with the 
ATCA at 9\,mm (Norfolk et al. in prep.). 
The available photometric data, as well as the predicted photospheric flux densities 
in the specific bands are given in Table~\ref{phottable}. The uncertainty of the predicted photospheric fluxes 
is estimated to be 5\%. 
The compiled SED together with the photosphere model are displayed in 
Figure~\ref{fig:sedplot}. While at wavelengths shortward of 25{\micron} the measured flux densities are 
consistent with the predicted photospheric contribution, in the far-IR and millimeter regime CP$-$72~2713 exhibits 
significant excess emission that arises from an extended circumstellar dust disk (Sect.~\ref{sec:obs}). To estimate 
the characteristic temperature ($T_{\rm d}$) and fractional luminosity ($f_{\rm d} = L_{\rm disk} / L_{\rm bol}$) 
of this disk, we fitted the excess emission by a 
single temperature modified blackbody model where the emissivity is 1 at
wavelengths shorter than $\lambda_0$ and varies as $(\lambda / \lambda_0)^{-\beta}$ 
at longer wavelengths. This model accounts for the general observational results that the emission of debris 
disks exhibits a fall at wavelengths that are large relative to the typical size of the emitting 
grains. Because of the sparse wavelength coverage, the $\lambda_0$ parameter 
could not be adequately constrained in our case, thus in the fitting we fixed its value to 100{\micron}.  
We adopted a Levenberg-Marquardt approach \citep{markwardt2009} in searching for the best-fitting model. 
During the fitting process we used an iterative way to compute and apply color corrections for our photometric
data. As a result we derived a dust temperature of 43$\pm$3\,K and a $\beta$ of 0.05$\pm$0.08. 
The low $\beta$ 
value indicates an excess spectrum that is almost identical to a pure blackbody and therefore 
changing the $\lambda_0$ parameter has no effect on our results. 
We note that though most debris disks show considerably 
steeper millimeter slopes \citep{gaspar2012,macgregor2016,marshall2017}, interestingly, 
the excess SED of AU\,Mic can also be well reproduced by a blackbody \citep{matthews2015}.
For the fractional luminosity we obtained 
$f_{\rm d} = 1.1\times10^{-3}$.
 
%%%%%%%% SED FIGURE

\subsubsection{The dust component}
Using our ALMA millimeter measurement we estimated the dust mass 
in grains of sizes smaller than a few millimeter
($M_{\rm d}$). 
Assuming optically thin dust emission that can be characterized by a single dust temperature, the dust mass is
\begin{equation}
M_{d} = \frac{{F_{\nu}} d^2}{ B_{\nu}(T_{\rm dust}) \kappa_{\nu}}, \label{eq:mdust}
\end{equation}
where $d$ is the distance, $B_{\nu} (T_{\rm dust})$ is the Planck
function at 1.33\,mm for the characteristic dust temperature, and $\kappa_{\nu}$ is the dust 
opacity at the given frequency. Dust particles with sizes $>$100{\micron}, whose radiation dominates the 
emission at this wavelength \citep{hughes2018}, are thought to behave like blackbodies. At a radial distance 
of 140\,au from CP$-$72~2713, a star with a luminosity of 0.18\,L$_\odot$ (Sect.~\ref{sec:stellarprops}), 
the temperature of 
such grains is $\sim$15\,K. By adopting $\kappa_{\nu}=$2.3\,cm$^2$\,g$^{-1}$ 
for the dust opacity \citep{andrews2013}, Eq.~\ref{eq:mdust} provides a dust mass 
estimate of 0.21$\pm$0.03\,M$_\oplus$. 
The quoted uncertainty accounts for the errors of the measured flux density and the distance 
of the system.
Note, however, that the potential systematic uncertainties of
the dust opacity can introduce even a factor of 2--3 difference in the estimate
 \citep[e.g.][]{miyake1993,ossenkopf1994} 
and the dust temperature may not be characterized by a single value.
These factors were not considered in our error calculations.  

\subsubsection{The gas component}
To derive an upper limit for the total CO gas mass using the measured $^{12}$CO 
integrated line flux upper limit, we need to estimate the fraction of CO 
molecules that populate the upper level of the 2--1 rotational transition. 
Collisions of CO molecules with different gas constituents, radiative processes, and 
UV pumping \citep{matra2015,matra2019} can all affect the population of the 
rotational levels. Based on the available data, it is difficult to judge the 
relative importance of these processes in the excitation. In our calculations 
we assumed that local thermodynamic equilibrium holds, i.e. that collisions
dominate the excitation processes and that gas emission is optically thin. 
Adopting gas temperatures between 10 and 45\,K we derived CO mass upper limits 
in the range of 6.8--9.4$\times$10$^{-6}$\,M$_\oplus$.

\subsection{Source confusion and contamination check} \label{sec:contamination}
So far in our analysis we have assumed that the observed excess emission solely 
originates from a circumstellar disk. Knowing the main properties of the source, 
at this point it is worth assessing the possibility of source confusion as well 
as potential influence of different kinds of contaminations. 

First we examine whether the excess could come from a background 
galaxy. 
Based on the recently derived double power-law model of the cumulative 
number counts of galaxies at millimeter wavelengths \citep{gonzalez-lopez2020}
we estimated a probability of $\sim$0.006 of the presence of a background galaxy brighter than 
 3.8\,mJy within the primary beam of the ALMA~7-m array. If we consider 
 only the interferometric synthesized beam, this value is reduced to $\sim$1.1$\times$10$^{-4}$.    
Besides the low probability of positional coincidence, there are other 
strong arguments against this scenario. We measured an angular size of $\sim$7\farcs5
for the emitting structure at 1.3\,mm. As interferometric observations show, galaxies are 
compact at (sub)millimetric wavelengths with a typical size of 0\farcs3$\pm$0\farcs1 
\citep[][and references therein]{franco2018}, i.e., they are significantly smaller 
than our source. The SED of the measured excess emission towards CP$-$72~2713 peaks at 
$\sim$120{\micron} and has a millimeter spectral index of $\sim$2. 
Although the SEDs of local (low-redshift) galaxies peak at a similar wavelength, their millimeter 
spectral slope is significantly steeper with typical indices between 3.5 and 
4.0 \citep{casey2014}. Moreover, such low-redshift objects represent only a small fraction 
of known millimeter galaxies \citep[][and references therein]{casey2014}.

Lensed galaxies deserve a special attention since
gravitational lensing effect can magnify distant background sources 
thereby increasing both their fluxes and angular sizes. 
At the spatial resolution we achieved in our ALMA observation, 
multiple images of lensed galaxies can mimic a source
having a more comparable size to that of our target than normal
(unlensed) dusty galaxies. However, despite their significant 
contribution to the bright-end population of galaxies at 
submm/mm wavelengths, lensed dusty star-forming galaxies
are rare objects with a sky density of 0.1--0.2\,deg$^{-2}$
\citep{casey2014,negrello2010,negrello2017,wardlow2013}. Thus 
the probability that such an object appears within the primary 
beam of our millimeter observation is less than 4$\times$10$^{-5}$.
The shape of their SEDs also deviates from what we measured towards 
CP$-$72~2713: the redshifts of lensed galaxies are typically $>$2 \citep{vieira2013}, 
their long wavelength emission peaks at $\gtrsim$250{\micron} \citep{negrello2017}.

Although it is very unlikely that all observed excess is coming 
from a background galaxy, owing to our coarse spatial resolution, 
positional coincidence with fainter extragalactic objects cannot be 
completely ruled out. 
Recent high spatial resolution millimeter images 
of several debris disks show compact sources -- most likely background galaxies -- co-located or located 
close to the dust ring \citep{chavez-dagostino2016,booth2017,marino2017,su2017,faramaz2019}. 
Such objects, if present, can contribute to the measured emission at far-IR and millimeter 
wavelengths and can modify the derived millimeter spectral index of our 
target. Higher resolution millimeter observations are needed to explore 
these possible contaminating sources and to assess their role, 
e.g., in the observed unusually shallow millimeter slope.

In our analysis the modelling of the stellar contribution to the measured emission was limited to the photospheric 
fluxes. However, at millimeter wavelengths low mass active stars can exhibit non-photospheric 
emission
\citep[e.g.,][]{liseau2015,macgregor2015}. Recent millimeter ALMA observations of AU\,Mic revealed a central 
point source at the star's position with a flux density $\sim$6$\times$ higher than predicted based on the pure 
photospheric model \citep{macgregor2013}. The observed excess was attributed either to an inner dust belt 
\citep{macgregor2013} or coronal emission from the star itself \citep{cranmer2013}. In the case of CP$-$72~2713 
the 1$\sigma$ measurement uncertainty (0.37\,mJy) of the 1.33\,mm observation is $\sim$80$\times$ higher than 
the predicted photospheric flux density. Supposing that the non-photospheric contribution at CP$-$72~2713 is not 
stronger than in the case of AU\,Mic \citep[based on their fractional X-ray luminosity, see][the coronal activity levels of the two stars are similar]{kiraga2012}
i.e. $\lesssim 6\times$ 
the photospheric emission, the measurement uncertainty is significantly higher than the uncertainty of the 
stellar flux prediction. For shorter periods, however, a higher level of activity cannot be excluded. 
As it was demonstrated in the cases of Proxima\,Cen and AU\,Mic, a strong flare can 
enhance a star's millimeter flux by orders of magnitudes for a few 
tens seconds \citep{macgregor2018,daley2019}. 
\citet{white2018} also found significant variability in the 9\,mm emission of 
the young M-type star, HD\,141569\,B, throughout its $\sim$1\,hr 
observation.
By examining the obtained ALMA data  
stream of CP$-$72~2713, we found no evidence for such flux enhancements.
Actually, this meets  our expectations: we identified only 
51 flare events in the 52~day long TESS light curves 
(Sect.~\ref{sec:stellaractivity}), and found that the star spends only 
$\sim$1.2\% of the time in flare phase in the studied period 
(in fact the peaks of the flares, which can substantially 
contribute to the millimeter emission, cover even much less time). 
Based on all of this, we believe that 
most of the measured millimeter flux is due to the dust disk.

\subsection{Star-disk alignment}
For the projected rotational velocity of the star, \citet{torres2006} obtained a value of 
$v\sin{i_*} = 7.5\pm1.2$\,km~s$^{-1}$, while \citet{weise2010} quoted $v\sin{i_*} = 6.6\pm1.4$\,km~s$^{-1}$
in their work. Computing the weighted mean of these $v\sin{i_*}$ measurements and
using the rotational period (in days) and the stellar radius derived above (in solar radii), 
following \citet{campbell1985} and \citet{greaves2014} we estimated the inclination of the stellar pole 
with respect to the line of sight as:
\begin{equation}
\sin{i_*} = 0.0198 \frac{P v\sin{i_*}}{R_*}. 
\end{equation}
This yields an inclination of $i_* = 42\pm7${\degr}, that is in good agreement with the disk 
inclination obtained from the analysis of our ALMA data (Sect.~\ref{sec:almaobs}). 
This finding coincides well with the conclusions made by \citet{watson2011}, \citet{kennedy2013} and 
\citet{greaves2014}: by examining larger debris disk samples these authors also 
found no evidence for significant misalignment between the stellar equator and the disk 
 in the studied systems.

%-----------------------------------------------------------------
% DISCUSSION
%-----------------------------------------------------------------

\section{Discussion} \label{sec:discussion}

\subsection{Evolutionary status of the disk}

A noteworthy fraction of members in the $\sim$24\,Myr old BPMG exhibit excess emission at infrared
wavelengths \citep{rebull2008,rm2014,moor2016}. While in most of these 
systems the observed excess is likely attributed to thermal emission of tenuous 
circumstellar debris disks, at least one member of this group, the late-type close binary 
V4046~Sgr harbors a long-lived gas-rich protoplanetary disk \citep[e.g.][]{rosenfeld2013}. 
Thus, the presence of a protoplanetary disk even at the age of the BPMG is not 
unseen. Therefore in the following we will examine the nature of the CP$-$72~2713 disk.

Observationally one of the most conspicuous differences between protoplanetary 
and debris disks is related to the amount of dust. While protoplanetary disks 
with their larger dust content are optically thick at most wavelengths, debris disks 
are optically thin across the spectrum.  This is manifested in the fractional 
luminosities, the threshold between the two classes is typically set to $f_d=0.01$, which is 
about 10$\times$ higher than that of CP$-$72~2713. 
The estimated dust mass of $\sim$0.2\,M$_\oplus$ in our target matches 
well those of other debris disks \citep{holland2017}. 
Recent large surveys 
of protoplanetary disks implied a positive correlation between disk radius and millimeter continuum luminosity 
\citep[e.g.][]{andrews2010,tripathi2017,hendler2020}. The relationship between  
the effective disk radii ($R_{\rm eff}$) and the millimeter continuum luminosity ($L_{\rm mm}$) 
was studied in five nearby star forming regions (Ophiuchus, Taurus-Auriga, Chameleon~I, Lupus, Upper Sco) by 
\citet{hendler2020}.
In their work $R_{\rm eff}$ was defined as the radius that encircles 68\% of the continuum emission, while 
$L_{\rm mm}$ was the Band~7 flux density scaled to 140\,pc. Using our SED model (Sect.~\ref{sec:diskprops}) 
we estimated a flux density of 8.4\,mJy at 890{\micron} for CP$-$72~2713, that results in $L_{\rm mm} \sim 6\times10^{-4}$\,Jy.
With its low luminosity and large size of the disk, CP$-$72~2713 deviates significantly from the observed relationships:
protoplanetary disks with $R_{\rm eff}> 100$\,au have at least two orders of magnitude higher $L_{\rm mm}$. 
The dust properties of our target seem inconsistent with a protoplanetary nature.

The mass of protoplanetary disks is dominated by their gas component, 
which is no longer present in significant quantities in most debris disks \citep{wyatt2015,kral2018}. 
We observed no CO emission towards CP$-$72~2713.
The obtained 3$\sigma$ upper limit on the $^{12}$CO (2--1) line luminosity is 
about three orders of magnitude lower than that of the protoplanetary disk around V4046~Sgr 
\citep{rosenfeld2013}. We note that $^{12}$CO in V4046~Sgr is optically thick,
suggesting an even higher contrast in the gas mass ratio between the two systems.
Even more indicatively this upper limit also remains six times below the measured 
$^{12}$CO (2--1) line luminosity of the disk of $\beta$\,Pic, whose 
gas and dust components are attributed to collisional erosion of larger bodies thus 
found to be of second generation \citep{kral2016,matra2017,cataldi2018}. 
\citet{torres2006} and \citet{gaidos2014} found the H${\rm \alpha}$ line in emission 
in CP$-$72~2713 with measured equivalent widths of 1.9\,{\AA} and 0.84\,{\AA}, respectively. 
Based on the criteria proposed by \citet{barrado2003} these values can be explained with chromospheric 
activity and do not indicate ongoing accretion and thus the presence of an inner gas-rich disk. 

These findings suggest that our target harbors a gas poor disk and, together with the results 
on the dust component, favors the classification of CP$-$72~2713 as a debris disk. 

Nevertheless, with its fractional luminosity of $1.1\times10^{-3}$ this disk 
is considered as a very dust-rich system among debris disks. By comparing this value 
with that of other cold debris disks with late-type host stars in the BPMG, 
we can find that it is about $3\times$ higher than that of the iconic disk of AU\,Mic 
\citep[$3.5\times10^{-4}$,][]{matthews2015} and matches well the fractional 
luminosity of the debris disk around AG\,Tri 
\citep[$10^{-3}$,][]{rm2014}. Beside CP$-$72~2713 we know 
only four debris systems within 40\,pc of the Sun that have well established cold dust 
component and exhibit fractional luminosity higher than 10$^{-3}$, namely $\beta$~Pic, HD\,61005, TWA\,7 and 
HD\,107146.

Despite its large dust content and proximity, the disk of CP$-$72 2713 
has remained undiscovered for a long time. This is largely due to 
the low luminosity of the host star: the system does not exhibit 
excess at mid-infrared wavelengths observed by the {\sl WISE} satellite 
and its far-infrared emission falls below the detection limit of all-sky surveys 
performed by the {\sl IRAS} and {\sl AKARI} satellites.

\subsection{Dust removal mechanisms} \label{sec:dustremoval}
In steady state, the production of dust particles by collisions in debris disks 
is balanced by grain removal processes through radiation and stellar wind forces.
In most known debris disks around main-sequence stars, the stellar radiation pressure 
plays an important role in controlling the dust dynamics by driving particles 
on more eccentric orbits than those of their parent bodies, and by expelling grains 
small enough from the system \citep[e.g.][]{krivov2010}. In lower luminosity 
K- and M-type stars, however, the radiation pressure is not strong enough 
to remove grains \citep{plavchan2005,reidemeister2011,schueppler2015}. 

\begin{figure}[h!]
    \centering
    \includegraphics[width=0.5\textwidth]{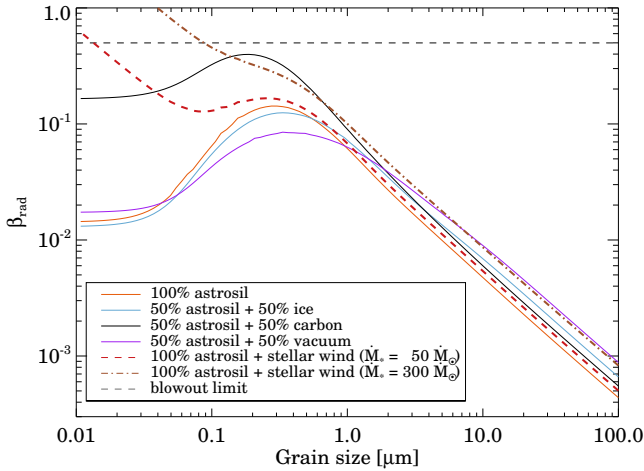}
    \caption{The $\beta_{\rm rad}$ ratio as a function of grain size for different dust compositions.
     Influence of the stellar wind pressure force is also explored for two different wind 
     mass loss rates. The blowout limit of $\beta$=0.5 is marked by a dashed horizontal line.}
    \label{fig:blowout}
\end{figure}

Figure~\ref{fig:blowout} 
shows the ratio of radiation to gravitational force, 
$\beta_{\rm rad} = \frac{F_{\rm rad}}{F_{\rm grav}} = \frac{3 L_* Q_{\rm pr}}{16 \pi G M_* c \rho s}$ \citep{burns1979}, 
for compact spherical dust grains with different $\rho$ densities and $s$ sizes in the CP$-$72~2713 system. 
We explored four different grain compositions, for which 
the $Q_{\rm pr}$ radiation  pressure efficiency was computed by means of Mie theory 
using the code developed by \citet{bohren1983}.   
Assuming parent bodies on a circular orbit the released grains become unbound if their 
$\beta_{\rm rad}$ is $>$0.5. As Fig.~\ref{fig:blowout} demonstrates, 
$\beta_{\rm rad}$ remains below this critical value for each 
tested dust composition, i.e. none of the particles are expelled by the 
radiation pressure in the CP$-$72~2713 system.
We note that as an active star, CP$-$72~2713 exhibits excess 
at UV wavelengths, shows strong coronal X-ray emission (Sect.~\ref{sec:stellaractivity}), 
and during flares its short wavelength emissions increases further.
Our stellar model does not account for these effects and thus 
we likely underestimate the strength of the radiation pressure. 
In the case of AU\,Mic whose UV, EUV, and X-ray emission is better 
constrained by observations than that of our target, 
\citet{augereau2006} found that this excess radiation is most important in the 
flare state when the $\beta_{\rm rad}$ value could be even a 
few times higher. 

The stellar wind mass loss of magnetically active late-type stars, as CP$-$72~2713 (Sect.~\ref{sec:stellaractivity}), 
is thought to significantly exceed the solar value \citep[e.g.][]{wargelin2001,wood2002}. 
In such systems wind forces can dominate the dust removal 
\citep{plavchan2005,strubbe2006,augereau2006}. Considering this corpuscular component
as well, the ratio of the radiation and wind forces to  gravitational force can be calculated 
following \citet{strubbe2006}: 
\begin{equation}
\beta_{\rm rad} = \frac{F_{\rm rad}+F_{\rm sw}}{F_{\rm grav}} = 
\frac{3 L_* P_{\rm swr}}{16 \pi G M_* c \rho s}, 
\end{equation}
where $P_{\rm swr} \equiv Q_{\rm pr} + Q_{\rm sw} \frac{ \dot{M}_{\rm *,wind} v_{\rm wind} c}{L_*}$. 
In this formula $Q_{\rm sw}$ is the  stellar  wind coupling coefficient, 
that gives the ratio of the effective to geometric cross section, $\dot{M}_{\rm *,wind}$ 
is the stellar wind mass loss rate, while $v_{\rm wind}$ is the stellar wind velocity. 
No data are available on the wind mass loss rate of our target in the literature. 
For the similarly active AU\,Mic, \citet{augereau2006} derived mass loss rates of 50 and 
300$\times$ higher than that of our Sun ($\dot{M}_{\rm \odot,wind} = 2 \times 10^{-14} 
{M}_{\rm \odot} {\rm yr}^{-1}$) by taking its quiescent phase 
and considering the influence of episodic flare activity, respectively. 
Using these mass loss rates 
and adopting $Q_{\rm sw} = 1$ and 
a wind velocity of 400\,km~s$^{-1}$ (that corresponds to the average wind velocity 
of the Sun) we recomputed the $\beta$ values for astrosilicate particles. 
As Figure~\ref{fig:blowout} shows, particles with  
sizes smaller than $\sim$0.013{\micron} in the  $\dot{M}_{\rm *,wind}= 50 \dot{M}_{\rm \odot,wind}$ scenario
and $\sim$0.09{\micron} in the  $\dot{M}_{\rm *,wind}= 300 \dot{M}_{\rm \odot,wind}$ scenario 
have a $\beta$ value $>$0.5. Thus by considering the possible wind pressure 
these small grains are blown out from the system.  

The Poynting-Robertson (P-R) drag as well as the stellar wind drag also affect the dynamics 
of dust particles: producing effective forces opposite to the direction of the 
orbital motion, these drags result in a gradual loss of the grains' angular momentum 
and lead them to spiral into the star. Assuming a particle originally having a circular 
orbit, the time necessary to reach the central star from a radius $R$ under the effect 
of P-R drag can be computed as 
$t_{\rm P-R} ({\rm yr}) = 400\frac{R({\rm au})^2}{M_*({\rm M_\odot})}\frac{1}{\beta_{\rm rad}}$ \citep{wyatt2005}.
Taking the disk radius of 140\,au (Sect.~\ref{sec:almaobs}), during the lifetime of the system ($\sim$24\,Myr) 
only grains with $\beta_{\rm rad} > 0.45$ could have spiraled into the star. 
Since the collision cascade probably started well after the birth of the system 
(see Sect.~\ref{stirring}), then the available time is even shorter 
and the threshold value for $\beta$ is even larger.
The stellar wind drag can work on a shorter timescale. 
According to \citet{plavchan2005} 
$\frac{t_{\rm P-R}}{t_{\rm sw}} = \frac{Q_{\rm P-R}}{Q_{\rm sw}} \frac{\dot{M}_{\rm *,wind}c^2}{L_*}$, 
thus by assuming $\frac{Q_{\rm P-R}}{Q_{\rm sw}} = 1$ and using the adopted 
$\dot{M}_{\rm *,wind}$ values of $50 \dot{M}_{\rm \odot,wind}$ or $300 \dot{M}_{\rm \odot,wind}$ 
the timescale of the wind drag could be $\sim$80 or $\sim$480$\times$ shorter, than the P-R timescale and 
can result in significant inward transport of larger dust particles as well. 
In the following we will investigate how these effects, considering the influence of collisions, 
can affect the spatial distribution of grains.

\subsection{Collisional evolution} \label{sec:ace}
We utilized the Analysis of Collisional Evolution 
\citep[ACE,][]{krivov2006,loehne2008} modelling code that can 
simulate the collisional evolution of the disk as well as the impact of transport 
processes on the dust distribution.
By solving numerically the 
Boltzmann-Smoluchowski kinetic equation this code enables to follow the evolution of a disk 
composed of subplanetary sized bodies down to micron-sized dust particles, considering the 
outcomes of collisions between these solids as well as the influence of the 
stellar gravity force and stellar radiative/corpuscular forces on them. 
ACE has already been applied successfully to explore the long term evolution 
of several debris disks \citep[e.g.][]{loehne2012,reidemeister2011,schueppler2015,geiler2019}.

In our simulations we used the stellar and disk parameters derived in this work. 
We adopted a stellar mass of 0.71\,$M_\odot$ and a luminosity of 0.18\,$M_\odot$, while 
for the stellar SED we used the ATLAS9 atmosphere model compiled in Sect.~\ref{sec:results}.
Based on the best fit model of the millimeter emission (Sect.~\ref{sec:almaobs}) 
 the parent planetesimal belt was assumed to be located between 126 and 154\,au corresponding 
 to a radial extent of 0.2$R$, with $R=140$\,au. 
We set a uniform surface density for this ring. {For the initial mass distribution of 
the solids we assumed a power-law with an index of -1.88 \citep{gaspar2012,krivov2018b}. 
The largest planetesimals were 
10$^{21}$g ($\sim$40km in radius; note that this size is not constrained, 
but was chosen to be sufficiently high to give a threshold for the collisional cascade), 
while the smallest particles were 10$^{-13}$g (sub-micron sized).
 The average initial orbital eccentricity of the
planetesimals was set to 0.03. Regarding the setup of the critical specific energy 
for disruption we followed that described by \citet{loehne2012}.
To explore the influence of the stellar wind we performed simulations with three different 
 wind mass loss rates of $\dot{M}_{\rm *,wind}= 0$, 
    $50\, \dot{M}_{\rm \odot,wind}$, and $300\, \dot{M}_{\rm \odot,wind}$.
The wind velocity was set to 400\,km~s$^{-1}$ in all three models.
We followed the disk evolution for 24\,Myr. 
By assuming pure astronomical silicate \citep{draine2003} as dust material, 
we calculated the SED of the simulated disks \citep[for details see][]{pawellek2019}. 
We found that disk models with a mass of 50\,$M_\oplus$ -- which include all solids up to 
the maximum size of 40\,km -- reproduce the measured SED at wavelengths $\leq1.3$\,mm 
well, although they cannot account for the unusually shallow millimeter SED, and thus 
substantially underestimate the observed flux at 9\,mm (see Figure~\ref{fig:sedplot} 
for an example). It is yet unclear what may be the reason for the measured low millimeter 
spectral index, but several different factors, including external contaminations 
(e.g. by galaxies, see Sect.~\ref{sec:contamination}), may play a role. 
To clarify these, and refine the model further, higher spatial resolution images 
are needed.

\begin{figure}[h!]
    \centering
    \includegraphics[width=0.5\textwidth]{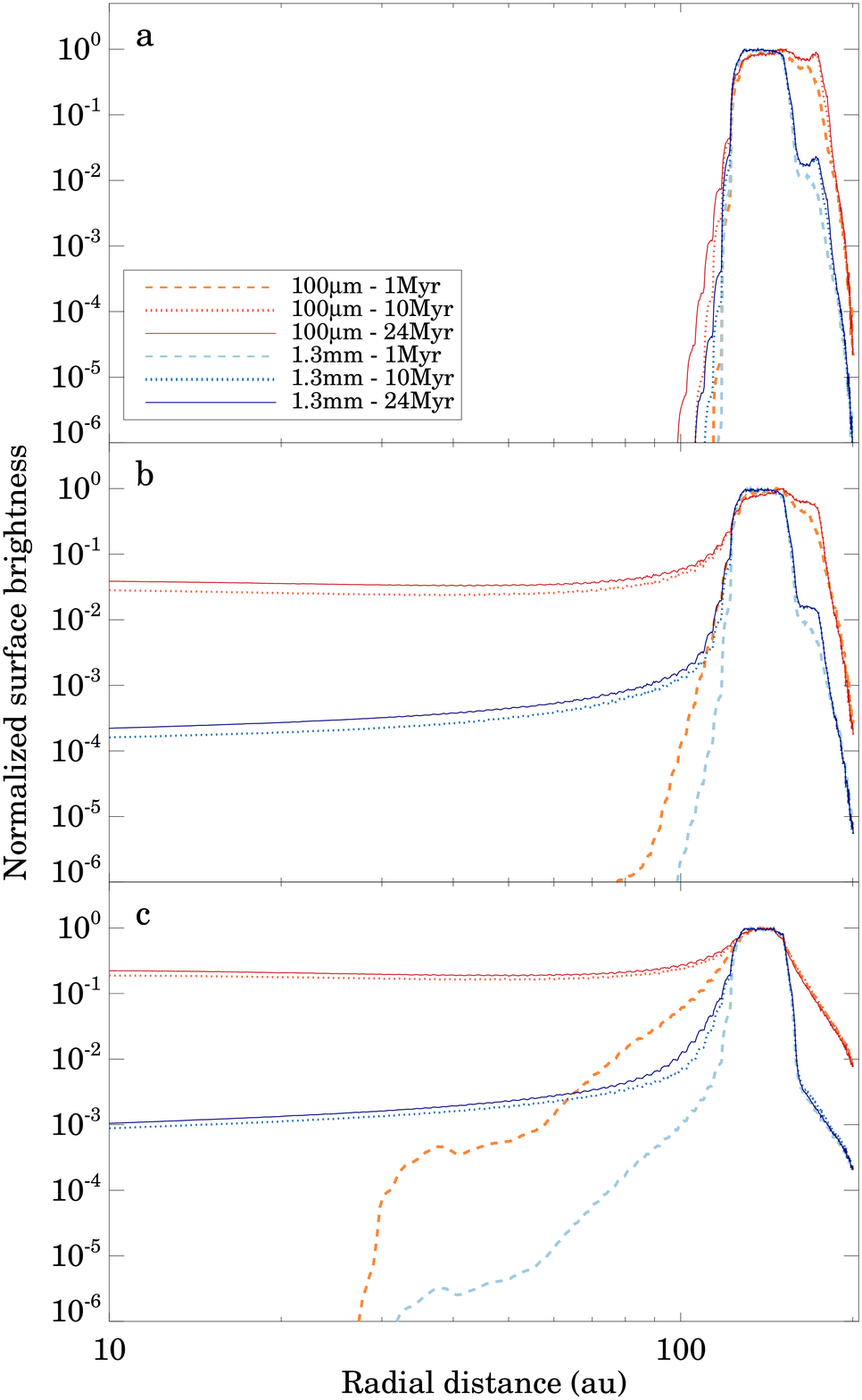}
    \caption{Normalized surface brightness at 100{\micron} and at 1.3\,mm 
    as a function of radial distance at three different evolution times. 
    From top to bottom, adopting different stellar wind mass loss rates of 
    (a) $\dot{M}_{\rm *,wind}= 0$, 
    (b) $\dot{M}_{\rm *,wind}= 50 \dot{M}_{\rm \odot,wind}$, 
    and (c) $\dot{M}_{\rm *,wind} = 300\, \dot{M}_{\rm \odot,wind}$.
    %$\dot{M}_{\rm *,wind}= 50 \dot{M}_{\rm \odot,wind}$ (b), and
    %$\dot{M}_{\rm *,wind}= 300 \dot{M}_{\rm \odot,wind}$ (c).  
    }
    \label{fig:ace}
\end{figure}

Using the 50\,$M_\oplus$ models we generated thermal maps at 100{\micron} and 
1.3\,mm.
Figure~\ref{fig:ace} displays 
normalized surface brightness profiles as a function of radial distance at these two 
wavelengths after 1, 10 and 24\,Myr evolution of the disk. These profiles were 
compiled as the azimuthal average of the thermal maps.
Assuming that the 
stellar wind is negligible (Fig.~\ref{fig:ace}a), 
we do not see significant differences between disks with different evolutionary 
states. This is consistent with the results of the above analytical calculations
(Sect.~\ref{sec:dustremoval}): the stellar radiation force itself is too weak 
to substantially modify grains' orbit. 
However, in models with stellar wind, the inward migration of dust particles 
becomes efficient due to the wind drag (Fig.~\ref{fig:ace}bc).  
At 1.3\,mm this has no significant impact on the surface brightness distribution. 
The thermal emission peaks at the planetesimal belt, the inner regions are two orders of 
magnitude fainter even in the $\dot{M}_{\rm *,wind}=300\, \dot{M}_{\rm \odot,wind}$ 
model. {This finding is consistent with the results obtained by \citet{pawellek2019}
for debris disks around late-type stars with stellar wind.}
At 100{\micron}, where smaller grains dominate the emission, the influence 
of the wind drag is more visible: though the birth ring is still the brightest part 
of the disk, especially in the model with the strongest wind the surface brightness 
inside the ring is only a few times lower.

\subsection{Possible signature of stellar wind drag?}

Interestingly, by analyzing the  PACS image we obtained a smaller 
disk radius ($99^{+22}_{-26}$\,au) than from the ALMA data (140$\pm$14\,au).
Though taking into account the uncertainties 
the difference between the two estimated radii is formally not significant, 
in the light of the ACE results it is worth examining whether it could 
be the effect of wind drag. 
We constructed simulated PACS images from the 100{\micron} thermal model maps 
by convolving them with our PSF model (Sect.~\ref{sec:pacsobs}).  
Then, using these simulated data we derived the disk peak radius by fitting 
the same Gaussian ring model as we applied in the case of the real data 
(Sect.~\ref{sec:pacsobs}). 
For most simulations the derived peak radii were around 140\,au, i.e. the surface brightness 
distribution is dominated by the parent belt.
However, for the most extreme model with $\dot{M}_{\rm *,wind}= 300\, \dot{M}_{\rm \odot,wind}$ 
at 24\,Myr we obtained a peak radius of 114\,au.

This result indicates that if the stellar wind is strong enough, 
then it may be able to shape the spatial distribution of grains so 
that the belt appears smaller at 100{\micron}.
We note, however, that taking into account that the formation of planetesimals 
and then the proper stirring of the disk (see Sect.~\ref{stirring}) needs time, the 
evolutionary time is likely substantially shorter than 24\,Myr. Thus, 
if the differences between the 100{\micron} and 1.3\,mm measurements are real and  
related to transport processes then this simulation suggests a 
$\dot{M}_{\rm *,wind}>300\, \dot{M}_{\rm \odot,wind}$ for this system.

Thanks to its large angular size and advantageous orientation the disk around 
CP$-$72~2713 could be an ideal target to investigate the influence of wind dominated 
transport in the future, however it needs high resolution multiwavelength 
observations that allow the study of the radial profile of the emitting regions.

\subsection{Location of the planetesimal belt}
Up to now about two dozens of debris disks have been spatially resolved at (sub)millimeter 
wavelengths, allowing to constrain the location of the outer planetesimal belts in these systems.
Using these data, \citet{matra2018} found a relationship between stellar luminosity $L_*$ and disk 
radius in the form of $R ({\rm au}) = 73^{+6}_{-6} L_*(L_\odot)^{0.19^{+0.04}_{-0.04}}$. This formula predicts 
a disk radius of 53$\pm$6\,au for CP$-$72~2713, which is significantly smaller than the measured 
size of 140\,au. 
 To further explore this aspect, in Fig.~\ref{fig:diskradii} we displayed 
 the outer radii\footnote{In the case of HD\,61005, \citet{macgregor2018b} modelled 
 the observed millimeter data using a two-component model composed by a planetesimal belt 
 and an outwardly extended halo. In our study we used the outer radius of the planetesimal belt 
 component (67\,au). 
 } of six debris disks ($\epsilon$~Eri, AU~Mic, HD\,61005, TWA\,7, HD\,92945, and HD\,107146) hosted by stars 
 with masses  
between 0.5 and 1.0~M$_\odot$ (similar to that of CP$-$72~2713) as a function of 
age. We have selected only objects that are younger than 1~Gyr and 
whose estimated radii are based on spatially resolved millimeter
observations \citep{booth2017,daley2019,macgregor2018b,matra2019,marino2018,marino2019} and thus 
thought to represent the outer edge of their planetesimal belts.
The age estimates of these systems are taken from the literature 
\citep{mamajek2008,bell2015,zuckerman2019,marino2019,williams2004}.
Similarly to CP$-$72~2713, these objects represent the most dust-rich known debris disks 
hosted by late-type stars in our neighbourhood. 
As Fig.~\ref{fig:diskradii} demonstrates, the disk radius of our target is comparable to 
those of HD\,92945 and HD\,107146, and is
substantially larger than those of the other three debris systems. This is especially remarkable when considering that 
in the case of CP$-$72~2713 we displayed the estimated peak radius of the belt instead 
of the outer radius (because of the limited angular resolution, the disk width was fixed 
in our modelling). Thus, CP$-$72~2713 is not just very dust-rich, but it is unusually extended.

\begin{figure}[h!]
\centering
\includegraphics[width=0.48\textwidth]{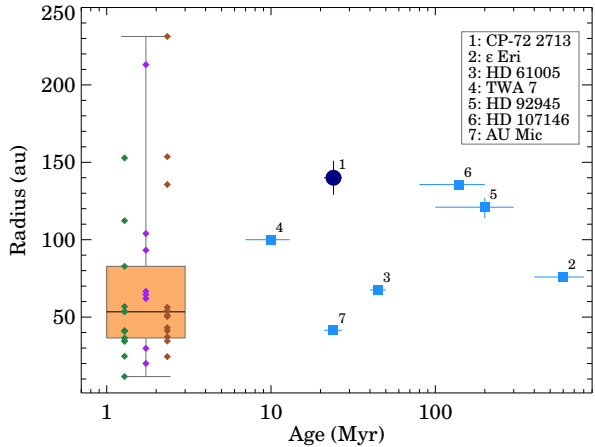}
\caption{Disk outer radius as a function of age. For CP$-$72~2713 the best-fit 
peak disk radius is displayed by a large dark blue circle.
Blue squares denote debris disks. Protoplanetary disks belonging to the 
Taurus, Lupus, and Chamaeleon~I star forming regions are marked by 
green, purple, and brown diamonds, respectively. The position of the 
points does not represent ages but serves to better distinguish the 
stars belonging to different regions. The shaded region represents 
the 25--75\% quartiles of the disk size distribution. The whiskers 
correspond to the 0--25\% and 75--100\% quartiles.
}
\label{fig:diskradii}
\end{figure}

Assuming that the planetesimal belt in the CP$-$72~2713 system is still close to its 
birthplace, we can expect that the predecessor protoplanetary disk
must have contained a concentration of a large amount of material at the same radius.
Thus it is interesting to compare the size of our disk with 
the measured outer radii of protoplanetary disks around similar mass (0.5--1.0\,M$_\odot$) 
young stars. Thanks to recent ALMA observations, there is a growing sample of protoplanetary 
disks resolved at millimeter wavelengths in the continuum. 
In Figure~\ref{fig:diskradii} we plotted the outer 
radii of 33 such protoplanetary disks belonging to the Lupus, Chamaeleon~I, 
and Taurus star forming regions. The ages of these groups 
range between 1 and 3~\,Myr. The disk size estimates of Lupus and 
Chamaeleon~I objects were taken from 
\citet{hendler2020}. We used their $R_{90}$ radii that are defined 
as radius containing 90\% of the millimeter continuum flux. 
For disks in Taurus we used the data from \citet{long2018} and 
\citet{long2019}, who derived $R_{90}$ and $R_{95}$ radii, respectively.
In Figure~\ref{fig:diskradii}, the shaded orange box shows the 
25--75\% quartiles of the size distribution, while the whiskers 
represent the 0--25\% and 75--100\% quartiles. The median disk 
radius is denoted by a horizontal black line.
We note that due to observational biases towards brighter objects the plotted 
disks likely represent the most massive/extended category of 
protoplanetary disks. Figure~\ref{fig:diskradii} shows that CP$-$72~2713 is 
not only the most extended debris disk but its radius is comparable to 
the largest protoplanetary disks in the studied 
stellar mass range.

\subsection{Stirring of the planetesimal belt} \label{stirring}
After the dispersal of the gas-rich protoplanetary disks the leftover planetesimals  
are thought to have nearly circular orbits, likely resulting in non-destructive,
low velocity collisions between them. 
To ignite an effective collisional cascade, that 
produces a continuous replenishment of dust particles observable at infrared and millimeter 
wavelengths, the planetesimal disk needs to be dynamically excited. The possible reasons of this 
dynamical excitation are still debated in the literature. The most commonly invoked explanations 
 require the existence of a giant planet or a stellar companion somewhere in the system 
 \citep[planetary or binary stirring,][]{mustill2009} or/and the presence of 
large planetesimals embedded in the belt \citep[self-stirring,][]{kenyon2008}. 
In our Solar System both self-stirring and planetary stirring may contribute to the 
excitation of the Kuiper-belt, resulting in a high excitation level there \citep{matthews2014}. 
However, our knowledge is limited on the origin of stirring in other 
debris disks \citep{moor2015,krivov2018}. All proposed stirring mechanisms need time 
to be activated in a given part of the disk. Using the model predictions 
thus we can examine their feasibility in the CP$-$72~2713 system.

\begin{figure}[h!]
\centering
\includegraphics[width=0.48\textwidth]{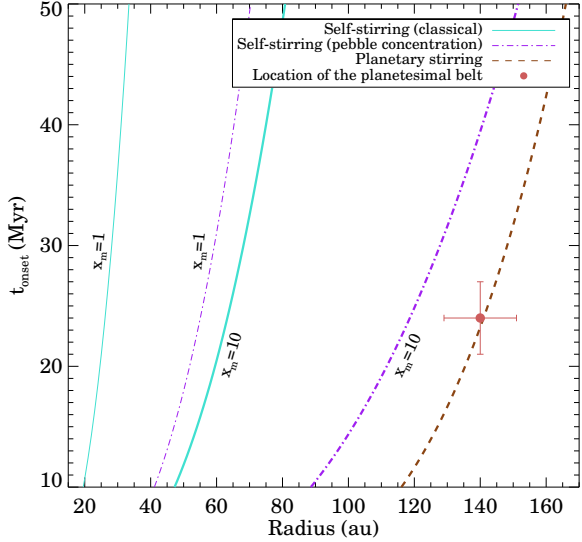}
\caption{Onset time of the stirring as a function of stellocentric radius for different 
stirring scenarios (Sect.~\ref{stirring}). The red dot corresponds to the observed property of the CP$-$72~2713 system. 
}
\label{fig:stirring}
\end{figure}

According to the classical version of the self-stirring model \citep{kenyon2008}, 
mutual low velocity collisions 
between smaller bodies in a dynamically cold region of the disk 
lead to the formation of gradually larger planetesimals. 
\citet{kenyon2008} found in their simulations that 
after reaching a radius of $\sim$1000\,km the largest planetesimals can stir 
their environment efficiently. It makes collisions between their smaller neighbours 
destructive and leads to the production of debris material through a collisional cascade.
Adopting an initial protoplanetary disk with a surface density distribution of 
$\Sigma(a) = 0.18 (M_*/M_{\sun}) x_{\rm m} (r/30{\rm au})^{-3/2} {\rm g~cm^{-2}},$ where 
$x_{\rm m}$ is a scaling factor (for the minimum-mass solar nebula $x_{\rm m} \equiv 1$), 
\citet{kenyon2008} provided an analytical formula for the formation time of 1000\,km-size planetesimals 
(the onset time of the stirring) 
at a stellocentric radius $r$: 
\begin{equation}
t_{\rm onset}^{\rm classical} = 145 x_{\rm m}^{-1.15} \bigg( \frac{r}{80\,{\rm au}} \bigg)^3 
   \bigg( \frac{2M_{\sun}}{M_*} \bigg)^{3/2}\rm Myr \label{t1000}.
\end{equation} 
Realistically, the $x_{\rm m}$ factor could not be higher than 10 \citep{mustill2009}. 
Taking a disk with the above-described radial surface density profile with $x_{\rm m} = 10$, 
and assuming that the location of the planetesimal belt at 140\,au in the CP$-$72~2713 system corresponds
 to the outer edge of the initial protoplanetary disk, the total initial mass of solids would have been 
 $\sim$1.2$\times$10$^3$\,M$_\oplus$. This hugely exceeds the measured dust masses 
 of protoplanetary disks around similar mass stars in the Lupus and Chameleon~I star forming regions 
 \citep{ansdell2016,pascucci2016}, further confirming the usage of $x_{\rm m} = 10$ as a strong upper 
 limit\footnote{We note that \citet{kenyon2008} derived 
Eq.~\ref{t1000} by fitting their model results for disks 
with $x_{\rm m}$ values between 1/3 and 3.}. In Figure~\ref{fig:stirring} we plot the onset time of 
the stirring at different radii 
 for $x_{\rm m} = 1$ and $x_{\rm m} = 10$ (solid cyan lines). A comparison with the location 
 of the planetesimal belt in the CP$-$72~2713 system (red circle), demonstrates that even 
 with $x_{\rm m} = 10$ the radius of the predicted stirring front is very far 
 from the region of the observed planetesimal belt at the age of the system.

The collisional coagulation model is not the sole feasible way for the formation of 
large planetesimals. Particle concentration models predict a significantly 
quicker formation of larger bodies even far from the star (at $r>$100\,au)
through gravitational collapse of locally concentrated pebbles \citep[][and references therein]{johansen2015,krivov2018}. 
According to these scenarios large planetesimals are already present 
in the disk by the time of the dissipation of the gas-rich protoplanetary material.
Due to the possibly smaller size of the largest bodies \citep[few hundreds kilometers,][]{simon2016,schafer2017} 
with respect 
to the classical model (where the 1000\,km sized 
planetesimals can stir the neighbouring disk promptly), the excitation of the surrounding smaller bodies 
takes more time than in the previous scenario. According to \citet{krivov2018}, 
in this scenario the onset time of effective self-stirring at a radius $r$ can be 
computed as 
\begin{equation}
\begin{aligned}
t_{\rm onset}^{\rm pebble} = \bigg( \frac{129}{x_{\rm m}} \bigg)
               \bigg( \frac{1}{\gamma} \bigg) 
	       \bigg( \frac{\rho}{1{\rm gcm^{-3}}}\bigg)^{-1} 
	       \bigg( \frac{v_{\rm frag}}{30{\rm ms^{-1}}} \bigg)^4 \\
               \bigg( \frac{S_{\rm max}}{200{\rm km}} \bigg)^{-3} 
	       \bigg( \frac{M_*}{M_\odot} \bigg)^{-3/2} 
	       \bigg( \frac{r}{100{\rm au}} \bigg)^3 ({\rm Myr}),
\end{aligned}
\end{equation}
where $\gamma$ is a parameter in eq.~20 in \citet{krivov2018}, $\rho$ is the bulk density of planetesimals, $v_{\rm frag}$ 
is the minimum collisional velocity needed for fragmentation, and $S_{\rm max}$ is   
the size of the largest planetesimals.
Figure~\ref{fig:stirring} shows -- by using the default values proposed by \citet{krivov2018} of 
$\gamma=1.5$, $\rho=1{\rm gcm^{-3}}$, $v_{\rm frag} = 30 {\rm ms^{-1}}$, and $S_{\rm max} = 200{\rm km}$ 
 -- that the outer edge of the possibly stirred region (purple dashed-dotted line) is close to the position of the 
 planetesimal belt in the CP$-$72~2713 system. Here we again adopted $x_{\rm m} = 10$.
 Considering the caveats of the model and the uncertainties of the fundamental model parameters
 \citep[][]{krivov2018}, it means that the excitation of the CP$-$72~2713 disk can be consistent
 with the pebble concentration stirring model. 
 For instance, by adopting somewhat larger maximum planetesimal sizes of $S_{\rm max}$=230\,km, 
 the model would predict exactly 24\,Myr for the onset of stirring at 140\,au, the location of the belt.

Dynamical excitation of a planetesimal disk can also be caused by a planet via its 
secular perturbation \citep{wyatt2005,mustill2009}. Similarly to the self-stirring scenario, 
the gravitational influence of a planet located within the planetesimal belt 
manifests in an outward propagating 
stirring front. Using the model developed by \citet{mustill2009} and assuming
a Saturn-mass planet with an orbital radius of 65\,au (roughly corresponding to the orbital radius 
of the outermost planet in the HR\,8799 system) and a moderate eccentricity of 0.1, we found 
that the perturbation induced by such a  planet could just reach the region 
of the belt by the age of the 
system (Fig.~\ref{fig:stirring}). The same result can be obtained with a smaller planetary mass 
if the semimajor axis 
and/or the eccentricity of the orbit is larger. Thus planetary stirring is a feasible model for the 
 CP$-$72~2713 system.
 
Finally, we cannot exclude the possibility that the planetesimal belt 
is pre-stirred, i.e. born stirred \citep{wyatt2008} because of 
some not yet specified physical mechanism. In this scenario the $t_{\rm onset}=0$ at 
all radii.

%%%%%%%%%%%%%%%%%%%%%%%%%%%%%%%%%%%%%%%%%%%%%%%%%%%%%%%%%%%

\section{Summary}

In this paper we present the results of our multiwavelength observations 
of a cold debris disk that we identified around CP$-$72~2713, a K7/M0-type 
member of the $\sim$24\,Myr old $\beta$\,Pic moving group. By analyzing the 
obtained SED we found that the excess spectrum is almost identical to a pure 
blackbody with a temperature of 43\,K and a fractional luminosity of 1.1$\times$10$^{-3}$.
No CO line emission was detected in the disk.
The derived dust fractional luminosity is prominently high, we know of only four 
other similarly dust-rich Kuiper-belt analogs within 40\,pc of the Sun.

{\sl Herschel} and ALMA images revealed that the observed emission at 100{\micron} and 
at 1.3\,mm is spatially 
resolved. Analysis of these data allowed us to explore the basic disk morphology.
Based on the derived disk orientation there is no evidence for any significant 
misaligment between the stellar equator and the disk plane, both are inferred 
to be inclined by $\sim$45{\degr}. Adopting a symmetric Gaussian radial surface 
brightness profile, our modelling yields disk radii of $99^{+22}_{-26}$\,au 
and 140$\pm$14\,au using the 100{\micron} and 1.3\,mm data, respectively.
By analyzing the possible dust removal mechanisms and the collisional evolution 
of the disk using the ACE code, we found that those large grains which 
dominate the observed millimeter emission are likely not moved significantly away 
from their birth location and thus trace well the parent 
planetesimal belt. Smaller grains emitting at 100{\micron}, however, might 
 spread inward if the stellar wind is strong enough, potentially explaining
 the smaller radius found with {\sl Herschel}.

A planetesimal belt radius of $\sim$140\,au is one of the largest
among known debris disks hosted by low mass stars.
In our neighbourhood we know of only a few protoplanetary disks around 0.5--1.0\,M$_\odot$ stars 
that contain a substantial amount of dust material at 140\,au, i.e. the location of the 
planetesimal belt in CP$-$72~2713. 
Supposing that this planetesimal belt was already at the same 
place when it was formed then it could be the offspring of a very extended 
protoplanetary disk.

We also examined the possible dynamical excitation of the planetesimal belt.
By applying different stirring models, we found that the presence of 
dust producing planetesimals at a radius of $\sim$140\,au is incompatible 
with the classical self-stirring scenario that assumes 
slow incremental growth of planetesimals. Stirring by a planet or self-stirring 
by planetesimals formed rapidly by pebble concentration, however 
can explain the dust production even in this extended, young disk.

AU\,Mic and CP$-$72 2713 both belong to the BPMG 
suggesting that they were born in the same star forming cloud 
at approximately the same time.
The two systems show further resemblances: their 
 late-type active central stars are surrounded by cold 
 debris disks. Therefore, considering its somewhat more massive host star 
 and more dust-rich and extended debris disk, CP$-$72~2713
appears to be a massive analog (a big sibling) of AU\,Mic. 
 Being the first representative of debris disks   
 orbiting a low mass star, AU\,Mic has been the target of many 
different studies \citep[e.g.][]{augereau2006,macgregor2013,boccaletti2015} and 
played a central role in understanding the 
physical mechanisms that can shape circumstellar material 
in such systems \citep{strubbe2006,schueppler2015}. The disk of CP$-$72~2713
has a similar angular size to that of AU\,Mic but it is closer to 
face-on, which makes it a more
favorable target to investigate the spatial structure of debris disks 
around low luminosity stars.

\bigskip

\acknowledgments

The  authors thank the anonymous referee for comments and 
suggestions which helped to improve the paper.
This paper makes use of the following ALMA data:
ADS/JAO.ALMA\#2017.2.00200.S. ALMA is a partnership of ESO
(representing its member states), NSF (USA) and NINS (Japan), together
with NRC (Canada) and NSC and ASIAA (Taiwan) and KASI (Republic of
Korea), in cooperation with the Republic of Chile. The Joint ALMA
Observatory is operated by ESO, AUI/NRAO and NAOJ. 
This work has made use of data from the European Space Agency (ESA) mission
{\it Gaia} (\url{https://www.cosmos.esa.int/gaia}), processed by the {\it Gaia}
Data Processing and Analysis Consortium (DPAC,
\url{https://www.cosmos.esa.int/web/gaia/dpac/consortium}). Funding for the DPAC
has been provided by national institutions, in particular the institutions
participating in the {\it Gaia} Multilateral Agreement.
This research has made use of the VizieR catalogue access tool, CDS,
 Strasbourg, France (DOI : 10.26093/cds/vizier). The original description 
 of the VizieR service was published in A\&AS 143, 23.
This work was supported by Hungarian NKFIH grant KH130526.

\vspace{5mm}
\facilities{Spitzer, Herschel, ALMA, TESS}

\software{CASA \citep{mcmullin2007}, uvmultifit \citep{marti-vidal2014}, mpfit \citep{markwardt2009}, 
MOPEX \citep[MOsaicking and Point source Extraction,][]{makovoz2005}, FLATW'RM code \citep{flatwrm}, 
MUFRAN \citep{mufran}, Stilism \citep{capitanio2017,lallement2018}}

\appendix

\section{Analysis of the TESS data} \label{appendix}

CP$-$72~2713 was observed in 2-minute (short) cadence by Camera 3 of {\sl TESS} spacecraft in Sectors 1 
(between 2018 July 25 and August 22) and 13 (between 2019 June 19 and July 18). In our analysis
we utilized the Simple Aperture Photometry (SAP) light curves that were  
downloaded from the MAST archive\footnote{\url{https://mast.stsci.edu}}.
We removed data points where the quality flags indicated problems using 
bit-wise AND operation with the binary mask 101010111111 as suggested by the TESS Data Product 
Overview\footnote{\url{https://outerspace.stsci.edu/display/TESS/2.0+-+Data+Product+Overview}}.
Zero-point of TESS magnitude scale (i.e. the magnitude - flux conversion factor) has been set 
using the GAIA DR2 RP magnitude \citep{brown2018}, 
by exploiting the almost complete overlap of the corresponding response functions at the 
far optical red and near-IR regimes (see fig.~1 in \citealt{ricker2015} and fig.~3 in \citealt{jordi2010}).
The obtained 
light curves show a clear periodic modulation likely due to rotation of 
stellar spots (Fig.~\ref{fig:tess}).

\begin{figure}[h!]
\begin{center}
\includegraphics[angle=0,scale=.40]{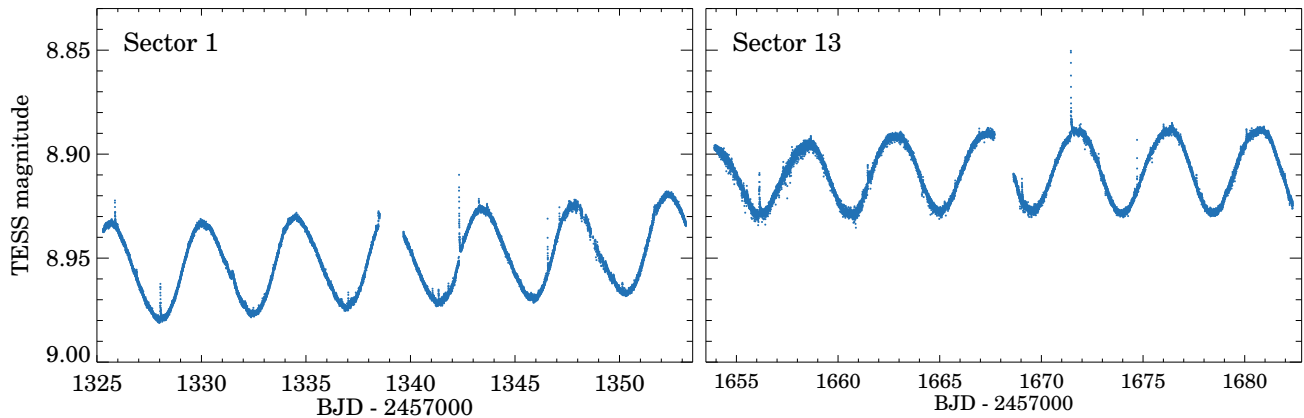}
\end{center}
\caption{{\sl TESS} short cadence light curves of CP$-$72~2713 measured in Sector\,1 (left) and in Sector\,13 (right).  
\label{fig:tess}}
\end{figure}

For the Fourier analysis of the short cadence light curves we subtracted a linear fit 
from the observations from each sector. Although these could be caused either by real 
intrinsic changes or instrumental artifacts, due to the distribution of the data points 
it is not possible to reliably fit variations on this time scale with Fourier components. 
The remaining data were analyzed using MUFRAN \citep{mufran}. The main feature in the 
Fourier-spectrum of the light curve (see Fig. \ref{fig:fourier}) is connected to the 
rotation ($P=4.437$ days), the other peaks -- an order of magnitude weaker -- are 
either present at the double frequency of this signal, or close to the main rotation 
period. These latter peaks could be a result of spots located at different latitudes 
showing a slightly different rotation rate due to differential rotation. A weak trend 
with the length of approximately the observing runs is also present.

Flares in the light curve were identified using the FLATW'RM code \citep{flatwrm}, that 
uses a machine-learning algorithm to give a robust model of the light curves in order 
to detect flare events and uses a voting system implemented to keep false positive 
detections to a minimum. We originally set a $3\sigma$ detection limit, that 
was lowered to $2\sigma$ as visual inspection of the results proved that there were 
no false positives identified. The minimum number of data points for a flare was set to 3, the 
degree of polynomials to describe light curves was set to 10. In the light curve 51 
events were identified: altogether 1.24\% of the data was considered part of a flare 
event. All the eruptions are relatively weak, each has an amplitude of $<0.04$ magnitudes.

\begin{figure}
    \centering
    \includegraphics[width=0.48\textwidth]{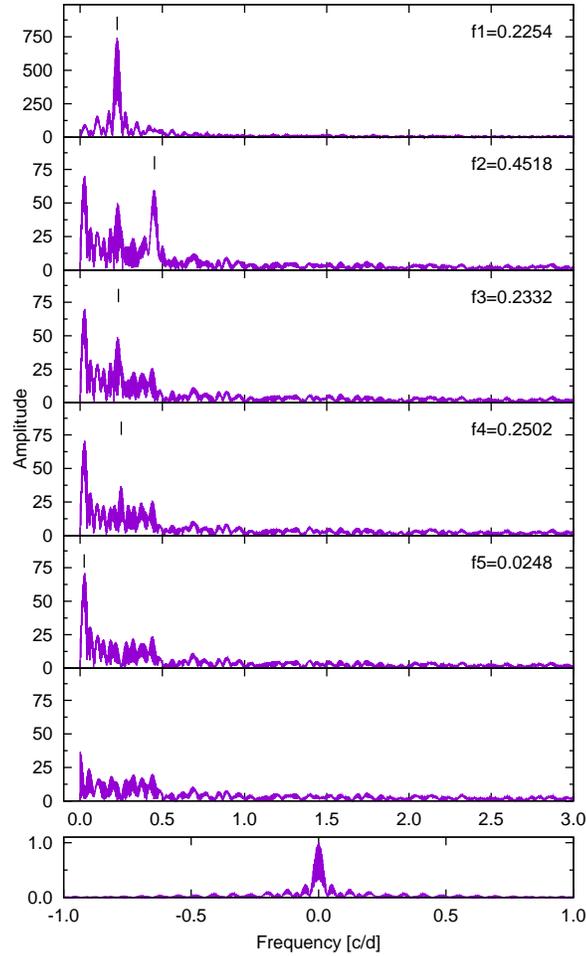}
    \caption{Fourier analysis of the TESS light curve. The upper five panels show the Fourier spectra of the light curve, after pre-whitening with the dominant frequency marked with a dash (note: the top plot has a different $y$ scale). The bottom two panels show the spectral window, and the residual light curve (note: largest flare is cropped), respectively.}
    \label{fig:fourier}
\end{figure}

\bibliographystyle{apj}
%\bibliography{adssample}

%% This command is needed to show the entire author+affilation list when
%% the collaboration and author truncation commands are used.  It has to
%% go at the end of the manuscript.
%\allauthors

%% Include this line if you are using the \added, \replaced, \deleted
%% commands to see a summary list of all changes at the end of the article.
%\listofchanges

\end{document}